\documentclass[structabstract]{aa}  
\usepackage{graphicx}
\usepackage{natbib}
\bibpunct{(}{)}{;}{a}{}{,}

\usepackage{txfonts}
\begin{document}
   \title{The protoplanetary disk of FT Tauri: \\
   multi-wavelength data analysis and modeling\thanks{Based on {\it Herschel} data. {\it Herschel} is an ESA space observatory with science instruments provided by European-led Principal Investigator consortia and with important participation from NASA.}}


   \author{A. Garufi
          \inst{1},
          L. Podio
          \inst{2, 3},
          I. Kamp
          \inst{4},
          F. M\'enard
          \inst{2, 5},
          S. Brittain
          \inst{6},
          C. Eiroa
          \inst{7},
          B. Montesinos
          \inst{8},
          \\
          M. Alonso-Mart\'inez
          \inst{7},
          W.F. Thi
          \inst{2},
          \and
          P. Woitke
          \inst{9,10,11}
          }

   \institute{Institute for Astronomy, ETH Z\"{u}rich, Wolfgang-Pauli-Strasse 27, CH-8093 Zurich, Switzerland\\
              \email{antonio.garufi@phys.ethz.ch}
              \and 
CNRS / UJF Grenoble 1, UMR 5274, Institut de Plan\'etologie et d'Astrophysique de Grenoble (IPAG), France
           \and
           INAF - Osservatorio Astrofisico di Arcetri, Largo E. Fermi 5, 50125, Florence, Italy
           \and
             Kapteyn Astronomical Institute, Postbus 800, 9700 AV Groningen, The Netherlands
          \and
          UMI-FCA, CNRS/INSU, France (UMI 3386), and Dept. de Astronom\'{\i}a, Universidad de Chile, Santiago, Chile
                   \and
Department of Physics \& Astronomy, 118 Kinard Laboratory, Clemson University, Clemson, SC 29634, USA
          \and
             Dpt. F\'isica Te\'orica, Facultad de Ciencias, Universidad Aut\'onoma de Madrid, Cantoblanco, 28049 Madrid, Spain
          \and
             Dpt. de Astrof\'isica, Centro de Astrobiolog\'ia, ESAC Campus, P.O. Box 78, E-28691 Villanueva de la Ca\~nada, Madrid, Spain
           \and
University of Vienna, Dept. of Astronomy, Turkenschanzstr. 17, A-1180 Vienna, Austria
            \and
UK Astronomy Technology Centre, Royal Observatory, Edinburgh, Blackford Hill, Edinburgh EH9 3HJ, UK
            \and
SUPA, School of Physics \& Astronomy, University of St. Andrews, North Haugh, St. Andrews KY16 9SS, UK
             }

   \date{Received ....; accepted ....}

 
  \abstract
   {Investigating the evolution of protoplanetary disks is crucial for our understanding of star and planet formation. Several theoretical and observational studies have been performed in the last decades to advance this knowledge. The launch of satellites operating at infrared wavelengths, such as the \textit{Spitzer} Space Telescope and the \textit{Herschel} Space Observatory, has provided important tools for the investigation of the properties of circumstellar disks.}
   {FT Tauri is a young star in the Taurus star forming region that was included in a number of spectroscopic and photometric surveys. We investigate the properties of the star, the circumstellar disk, and the accretion/ejection processes and propose a consistent gas and dust model also as a reference for future observational studies.}
   {We performed a multi-wavelength data analysis to derive the basic stellar and disk properties, as well as mass accretion/outflow rate from TNG/DOLoRes, WHT/LIRIS, NOT/NOTCam, Keck/NIRSpec, and {\it Herschel}/PACS spectra. From the literature, we compiled a complete Spectral Energy Distribution. We then performed detailed disk modeling using the \textit{MCFOST} and \textit{ProDiMo} codes. Multi-wavelength spectroscopic and photometric measurements were compared with the reddened predictions of the codes in order to constrain the disk properties.}
  {We determine the stellar mass ($\rm \sim 0.3 \ M_{\odot}$), luminosity ($\rm \sim 0.35 \ L_{\odot}$) and age ($\rm \sim 1.6 \ Myr$), as well as the visual extinction of the system (1.8 mag). We estimate the mass accretion rate ($\sim 3 \cdot 10^{-8}\ \rm M_{\odot}/yr$) {to be} within the range of accreting objects in Taurus. The evolutionary state and the geometric properties of the disk are also constrained. The radial extent (0.05 to 200 AU), flaring angle (power-law with exponent $=1.15$), and mass ($0.02 \ {\rm M_{\odot}}$) of the circumstellar disk are typical of a young {primordial} disk. This object can serve as a benchmark for primordial disks with significant mass accretion rate, {high gas content} and typical size.}
   {}

   \keywords{stars: pre-main sequence --
                planetary systems: protoplanetary disks --
                accretion, accretion disks --
                ISM: individual object: FT Tau
               }

\authorrunning{Garufi et al.}

\titlerunning{The protoplanetary disk of FT Tauri}

   \maketitle
%

\section{Introduction} \label{Intro}

Protoplanetary disks are the birthplaces of planets and the study of their physical and chemical structure can help us understand planet formation. By studying a large sample of young protoplanetary disks (class II) in detail, we may be able to assess the variety in disk structure and to match that to the ever growing diversity in exoplanetary systems architecture.

The disk evolution can be observationally constrained by studying the Spectral Energy Distribution (SED) of Young Stellar Objects (YSOs) in different evolutionary stages. In particular, the mass accretion rate can be estimated from the excess in the UV and optical spectra and from emission lines that are thought to form in the magnetospheric accretion process (Basri \& Bertout 1989, Edwards et al.\ 1994, Hartmann, Hewett, \& Calvet 1994).  

Also geometrical properties of circumstellar disks change with time. In the absence of spatially resolved images, the geometry has to be constrained from {infrared (IR)} and millimetric photometry. A flaring geometry is a natural explanation for the strong {far-infrared (FIR)} flux shown by most sources (Kenyon \& Hartmann 1987). Grain-grain collisions result in dust grain growth and, on timescales of $10^4-10^6$ years, grains are thought to settle to the mid-plane leaving the gas at the disk surface exposed to direct stellar radiation (Bouwman et al.\ 2008). This evolution is reflected in decreasing {mid-infrared (MIR)} fluxes and in a widening and flattening of the 10 $\mu$m and 18 $\mu$m silicate features (e.g.\ Furlan et al.\ 2006, Fang et al.\ 2009).

The diagnostic of gas emission lines from the disk is another pivotal tool for the study of the disk structure. CO and OH lines are commonly detected in the {near-infrared (NIR)} spectra of protoplanetary disks. In particular, the {CO fundamental ($\nu =1-0$) lines in the M band are excellent tracers of the temperature and density structure inside a few AU, the terrestrial planet-forming regions, because of their sensitivity to low column densities of gas at temperatures of a few 1000 K (Najita et al.\ 2000). On the other hand, the frequently observed FIR [O \textsc{i}] 63 $\mu$m line (Dent et al.\ 2013)} is believed to mostly originate in the colder outer regions, between 30 and 100 AU (Kamp et al.\ 2010). Its flux can be used in combination with other lines as an indicator of disk gas mass.            

An ever growing number of datasets is becoming available for circumstellar disks. Nevertheless, many studies still focus on the interpretation of a single dataset even in the framework of detailed disk modeling. However, full characterization of stellar and circumstellar properties of single objects is often reached only by employing dust and gas diagnostic measurements (photometry and line emission) that cover the entire extent of a protoplanetary disk. In this paper, we aim to investigate the geometrical and chemical properties of the disk around the T Tauri Star (TTS) FT Tauri, by means of a multi-wavelength dataset and consistent dust and gas modeling.

Even though the Taurus star forming region and its members are well studied (e.g.\ Kenyon et al.\ 1994, Gullbring et al.\ 1998, Luhman et al.\ 2010, Rebull et al.\ 2010), a comprehensive characterization has been restricted so far to either extremely bright (large) disks (e.g.\ DM Tau, Guilloteau \& Dutrey 1994) or exceptional objects (e.g.\ LkCa15, van Zadelhoff et al.\ 2001). FT Tau is located in the South of the Barnard 215 dark cloud and is surrounded by extended emission (see Sloan Digital Sky Survey optical image from Finkbeiner et al.\ 2004). This source is far from most of the known Taurus members (see extinction map of Taurus from Dobashi et al.\ 2005). No X-ray emission was detected at the optical position of the star (Neuh\"{a}user et al.\ 1995). FT Tau was included in a number of photometric and spectroscopic surveys at different wavelengths. {The main stellar and disk properties from the literature are shown in Table \ref{Properties}.} With more sensitive interferometers such as SMA and IRAM/PdBI, FT Tau can serve as an excellent target for more detailed astrochemical studies.

The multi-wavelength data used in this work and its data reduction is presented in Sect.\,\ref{Observations} and a detailed analysis thereof in Sect.\,\ref{Results}. The result of this analysis are then further used in Sect.\,\ref{Modeling} to build consistent dust and gas models (MCFOST and \textit{ProDiMo}) for FT Tau. Sect.\,\ref{Discussion} then discusses the results {and the source variability}.

\begin{table}
      \caption[]{Properties of FT Tauri estimated in previous works.}
         \label{Properties}
     $$ 
         \begin{tabular}{lcccc}
            \hline
            \hline
            \noalign{\smallskip}
            Coordinates (J2000): \\
            Right ascension & $\alpha$ & 04$^{\rm{h}}$ 23$^{\rm{m}}$ 39$^{\rm{s}}$.19 $^{\mathrm{\ a}}$ \\
            Declination & $\delta$ & +24$^{\circ}$ 56$'$ 14$''$.11 $^{\mathrm{\ a}}$ \\
            \hline
            Proper motion: \\
            Right ascension & $\mu _{\alpha}$ & +6.3 $\pm$ 3.3 mas yr$^{-1}$ $^{\mathrm{\ b}}$ \\
            Declination & $\mu _{\delta}$ & -15.3 $\pm$ 3.3 mas yr$^{-1}$ $^{\mathrm{\ b}}$ \\
            \hline
            Stellar properties: \\
            Visual extinction & $A_V$ & 3.8 $^{\mathrm{\ c}}$ \\
            Spectral type & & M3e $^{\mathrm{\ c}}$ \\
            Luminosity & $L_*$ & 0.63 L$_{\odot}$ $^{\mathrm{\ c}}$\\
            \hline
            Disk properties: \\
            Gas mass & $M\rm_{d}$ & $\rm 0.05 \pm 0.03 \ M_{\odot}$ $^{\mathrm{\ d}}$ \\
            Outer radius & $R_{\rm out}$ & $50^{+950} _{-25} \ \rm{AU}$ $^{\mathrm{\ d}}$ \\
            \noalign{\smallskip}
            \hline
            \noalign{\smallskip}
         \end{tabular}
     $$ 
$^{\mathrm{\ a}}$ Cutri et al.\ 2003; $^{\mathrm{\ b}}$ Luhman et al.\ 2009; $^{\mathrm{\ c}}$ Rebull et al.\ 2010;  $^{\mathrm{\ d}}$ Andrews \& Williams 2007.

   \end{table}


\section{Observations and data reduction} \label{Observations}
The data analysed in this paper consist of spectroscopy at optical, NIR, and FIR wavelengths from the Telescopio Nazionale Galileo {(TNG)}, the William Herschel Telescope {(WHT)}, the Nordic Optical Telescope {(NOT)}, the Keck Observatory, and the {\it Herschel} Space Observatory. {The instrumental settings for the spectroscopic observations are presented in detail in the following sections and summarized in Table \ref{Settings}}. Additional data were retrieved from the literature and consist of MIR spectroscopy from {\it Spitzer} Space Telescope and of photometry from optical to radio wavelengths {(see Table \ref{Photometry} and Fig.\,\ref{Spectra})}.  

\begin{table*}
      \caption[]{Instrumental settings for the spectroscopic observations of FT Tau.}
         \label{Settings}
     $$ 
         \begin{tabular}{lcccccc}
            \hline
            \hline
            \noalign{\smallskip}
            Observation date & Instrument & Slit width ('') & Spectral range ($\mu$m) & Spectral resolution (km/s) & Integration time (s) \\
            \hline
            \noalign{\smallskip}
            2009-11-26 & TNG/DOLORES & 1 & 0.475 - 0.670 & 196 & 1400, 600, 200 \\
            2009-12-04 & NOT/NOTCAM & 1 & 1.95 - 2.37 & 200 & 48 \\
            2009-12-09 & WHT/LIRIS & 0.75 & 1.17 - 1.35 & 135 & 120 \\
            2008-12-10 & Keck/NIRSPEC & 0.43 & 4.42 - 5.53 & 12 & 960, 1200 \\
            2010-03-26 & {\it Herschel}/PACS & Integral field & 62.95 - 63.40 & 88 & 1152 \\
            \noalign{\smallskip}
            \hline
            \end{tabular}
     $$ 

   \end{table*}

\subsection{TNG/DOLORES observations} \label{Observations_Tng}
We present spectroscopic data of FT Tau obtained using the DOLORES spectrograph, the Device Optimized for the LOw RESolution (Oliva 2004) mounted on the {TNG} (La Palma Observatory). The observations were performed in November 2009 with the VHR-V grism ($\lambda / \Delta \lambda = 1527$ for a slit width of 1''), covering the wavelength range from 4752 $\rm \AA$ to 6698 $\rm \AA$. 

Three observations (exposure times of 1400, 600, and 200 s) were performed in seeing-limited conditions (FWHM $\simeq$ 1.25''). The DOLORES spectra were reduced using the IRAF software package. The 2-D spectra were flat-fielded, sky-subtracted and wavelength calibrated using the Argon arc lamp. Then, the 1-D spectrum was extracted by integrating over the source spatial profile and corrected for telluric absorption features. No photometric standard observations were taken. Thus, the flux calibration was obtained from previous photometry in the V band (see Table \ref{Photometry}). {Figure \ref{Spectra}a shows that the V band photometry agrees well with the neighboring SDSS photometry.} This approach is not taking into account the source variability, which is further discussed in Sect.\,\ref{Variability}.

\subsection{WHT/LIRIS observations}
A J-band (1.18 - 1.40 $\mu$m) spectrum was taken in December 2009 with the LIRIS, Long-slit Intermediate Resolution Infrared Spectrograph, at the {WHT} (La Palma Observatory). The spectrum was acquired with a 0.75\arcsec slit width and the LIRIS hrj grism providing a spectral resolution of $\lambda / \Delta \lambda = 2200$. The exposure time was 120 seconds. The data reduction was carried out using IRAF. After flat-field correction and sky subtraction, an Argon lamp was used {for wavelength calibration of} the 1-D extracted spectrum. The flux calibration was performed by means of the 2MASS photometry (see Table \ref{Photometry} {and Fig.\,\ref{Spectra}b}).

\subsection{NOT/NOTCAM observations}
A K-band spectrum of FT Tau was taken with the NOTCAM, the Nordic Optical Telescope CAMera (Aspin 1999) of the {NOT} (La Palma Observatory). These observations were acquired in December 2009 by using the K grism ($\lambda / \Delta \lambda = 1500$ for a slit width of 1''), operating in the NIR K-band (1.95 - 2.37 $\mu$m), {and nodding} the slit between two positions. 

The observation was obtained in seeing-limited conditions (FWHM $\simeq$ 0.9'') with an exposure time of 48 s. The data reduction of the spectrum was performed using the IRAF software. The spectrum was background-subtracted and flat-fielded. The 1-D spectrum was extracted and wavelength calibrated through Argon arc lamp. Since no photometric standard observations were available, the flux calibration of these data was performed exploiting the K$\rm _S$ photometry from 2MASS (see Table \ref{Photometry} {and Fig.\,\ref{Spectra}c}).

\subsection{Keck/NIRSPEC observations} \label{Observations_Keck}
An M-band high-resolution spectrum of FT Tau was taken with the NIRSPEC, the NIR SPECtrograph (McLean et al.\ 1998) on the W.M. Keck Observatory. The spectra were obtained in December 2008 using the M-Wide filter and 0.43'' slit providing a resolution of $\lambda / \Delta \lambda = 25,000$. The spectra span 4.67 - 5.05 $\mu$m. FT Tau was observed for 16 and 20 minutes in successive exposures. 

Because of the thermal background in the M-band, the data were observed in an ABBA sequence where the telescope was nodded 12'' between the A and B positions. Each frame was flat fielded and scrubbed for hot pixels and cosmic ray hits. The observations were combined as (A-B-B+A)/2 in order to cancel the sky emission. The data were rectified in the spatial direction by fitting a polynomial to the point spread function (PSF) of the star in each column. The data were rectified in the spectral direction by fitting a sky emission model generated by the Spectral Synthesis Program (Kunde \& Maguire 1974) to each row and interpolating the wavelength solution to each row to the row in the middle of the detector. 

\begin{figure*}
   \centering
   \includegraphics[width=7cm]{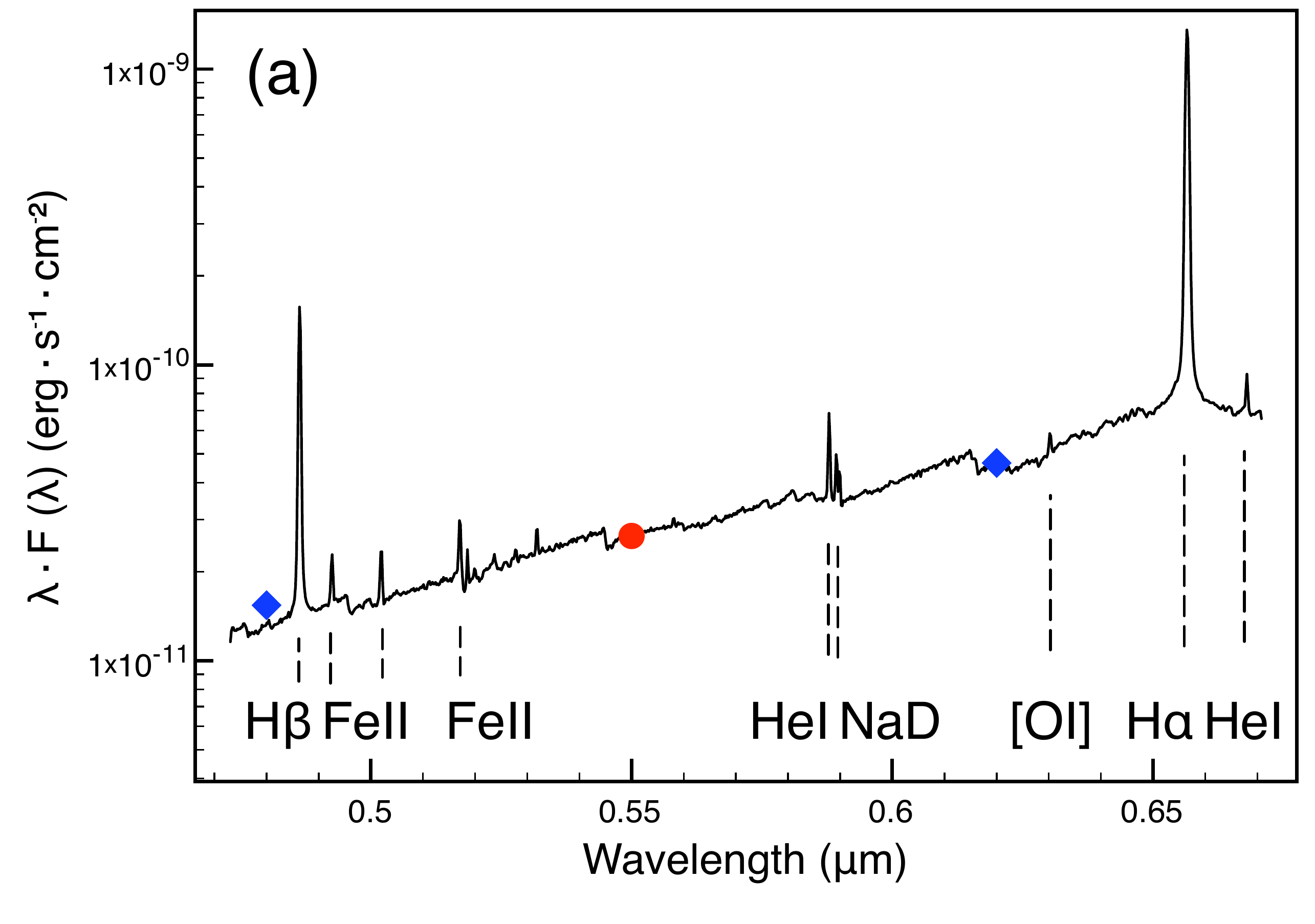}
   \includegraphics[width=7cm]{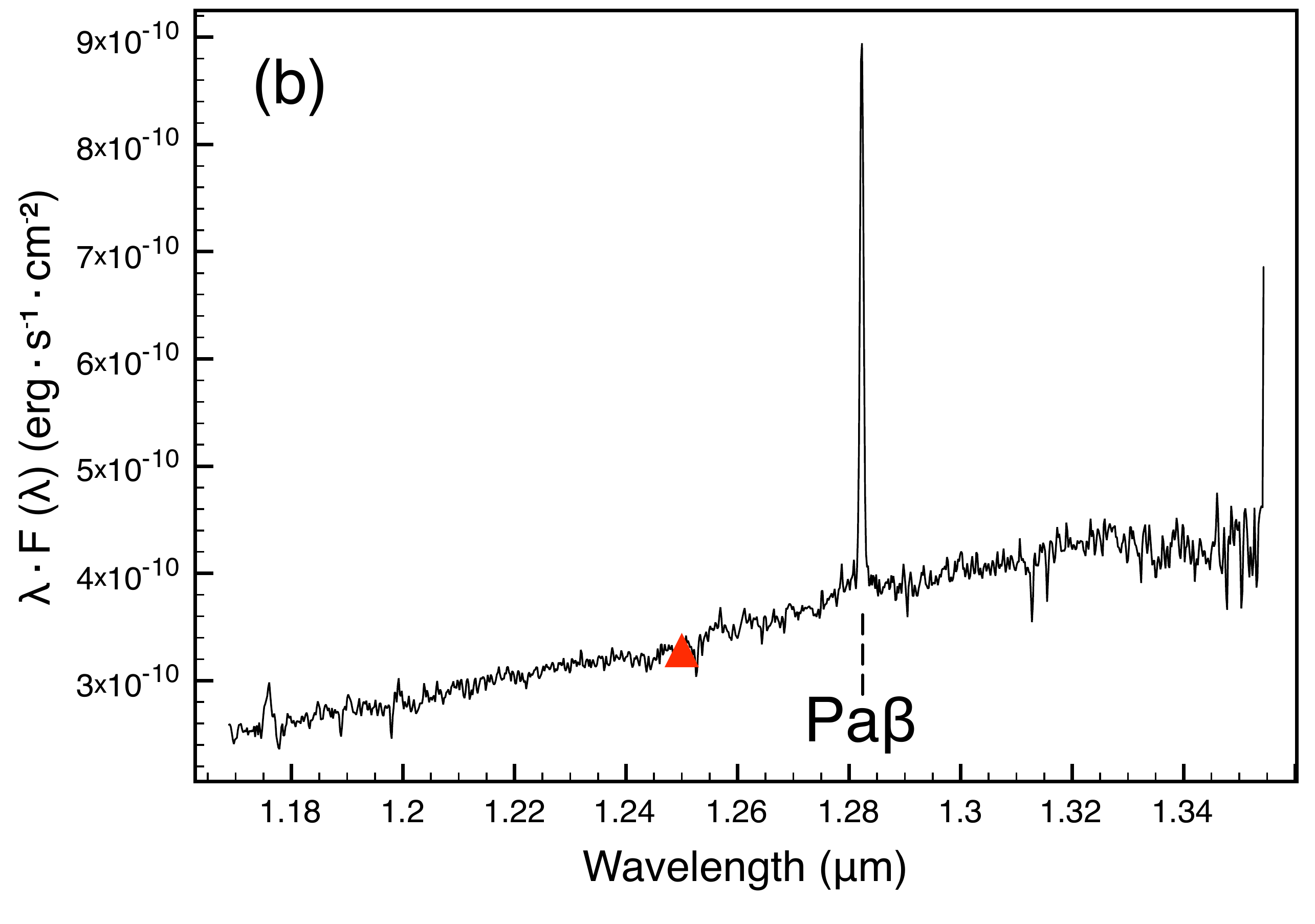}
   \includegraphics[width=7cm]{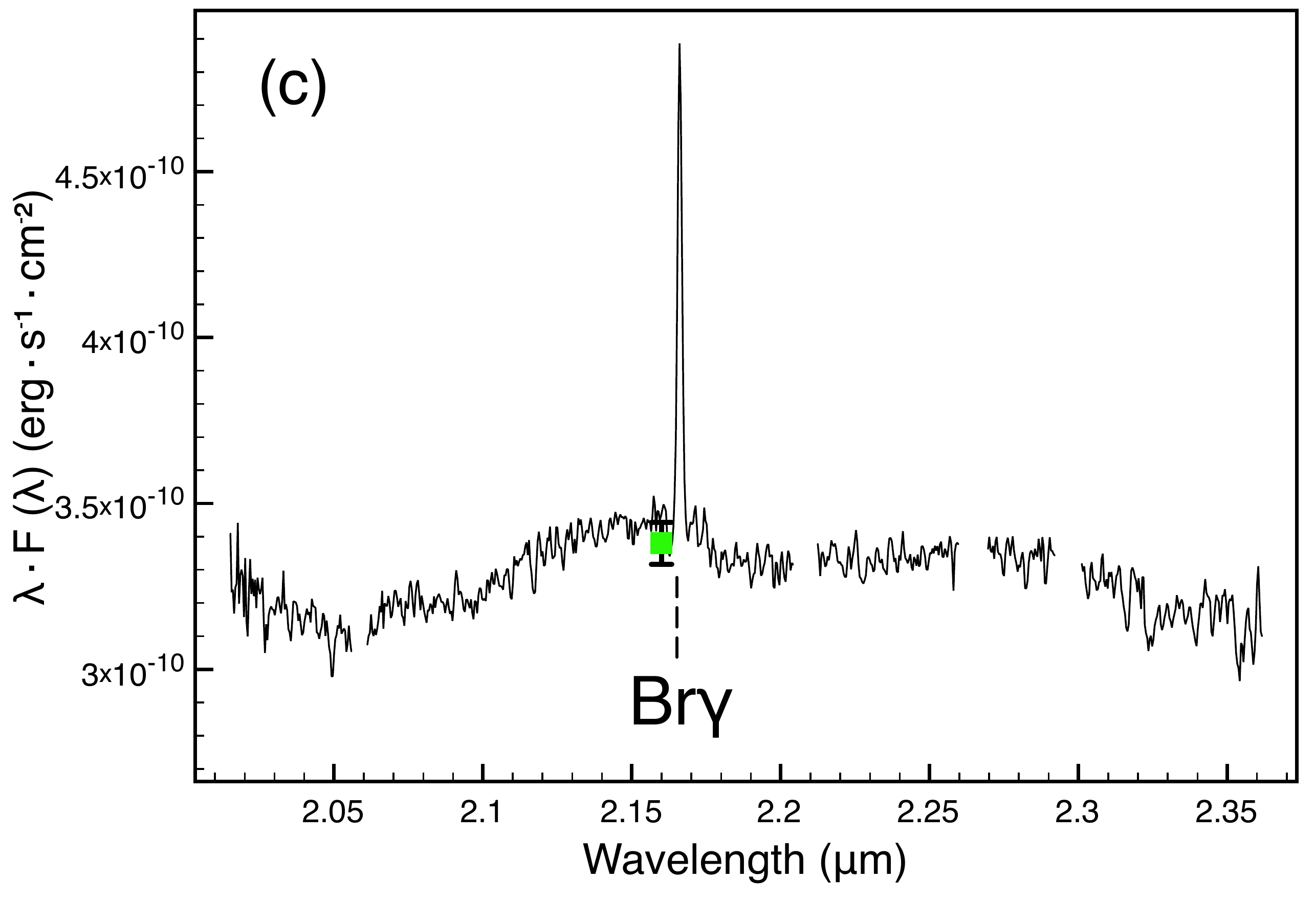}
   \includegraphics[width=7cm]{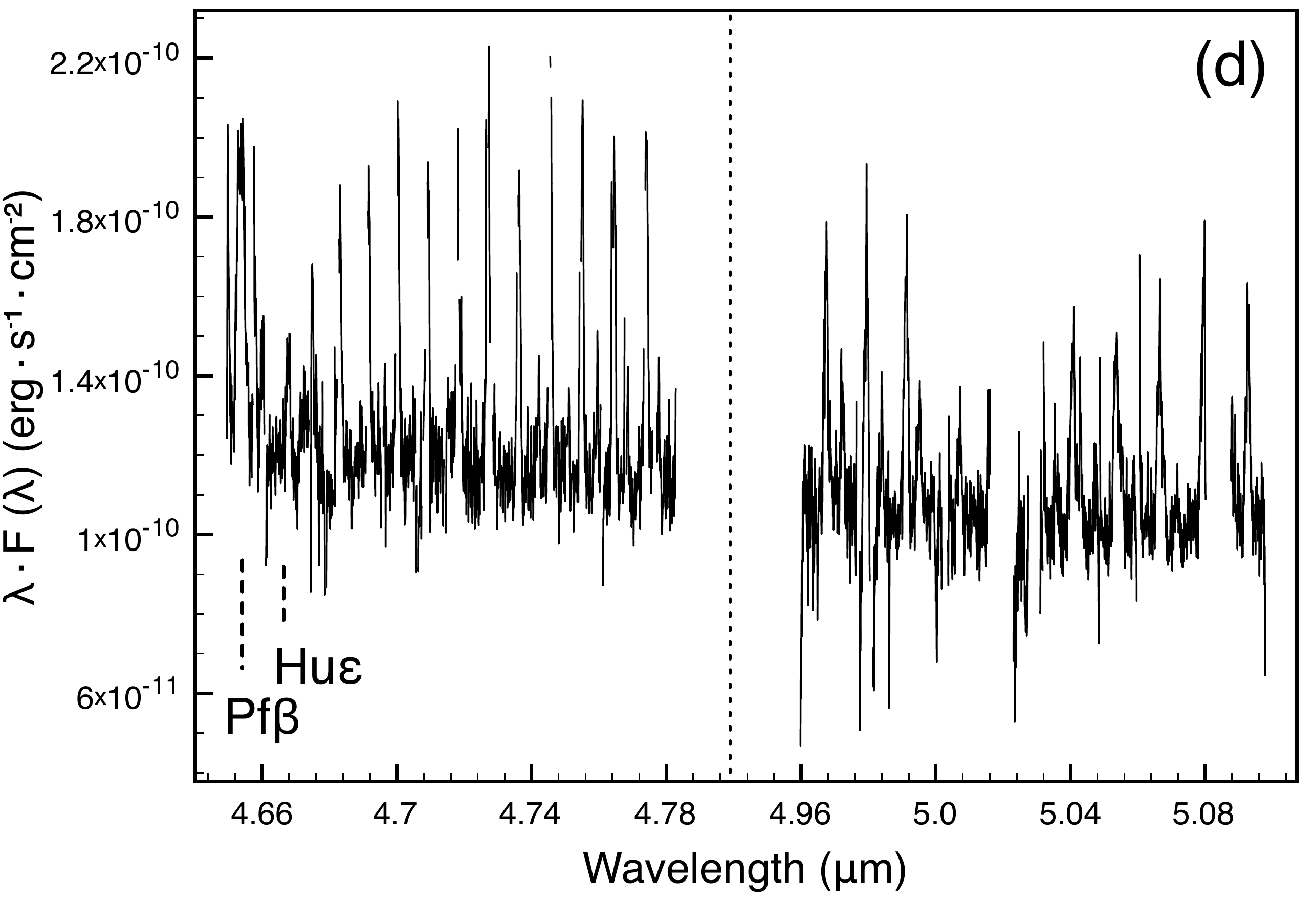}
   \includegraphics[width=7cm]{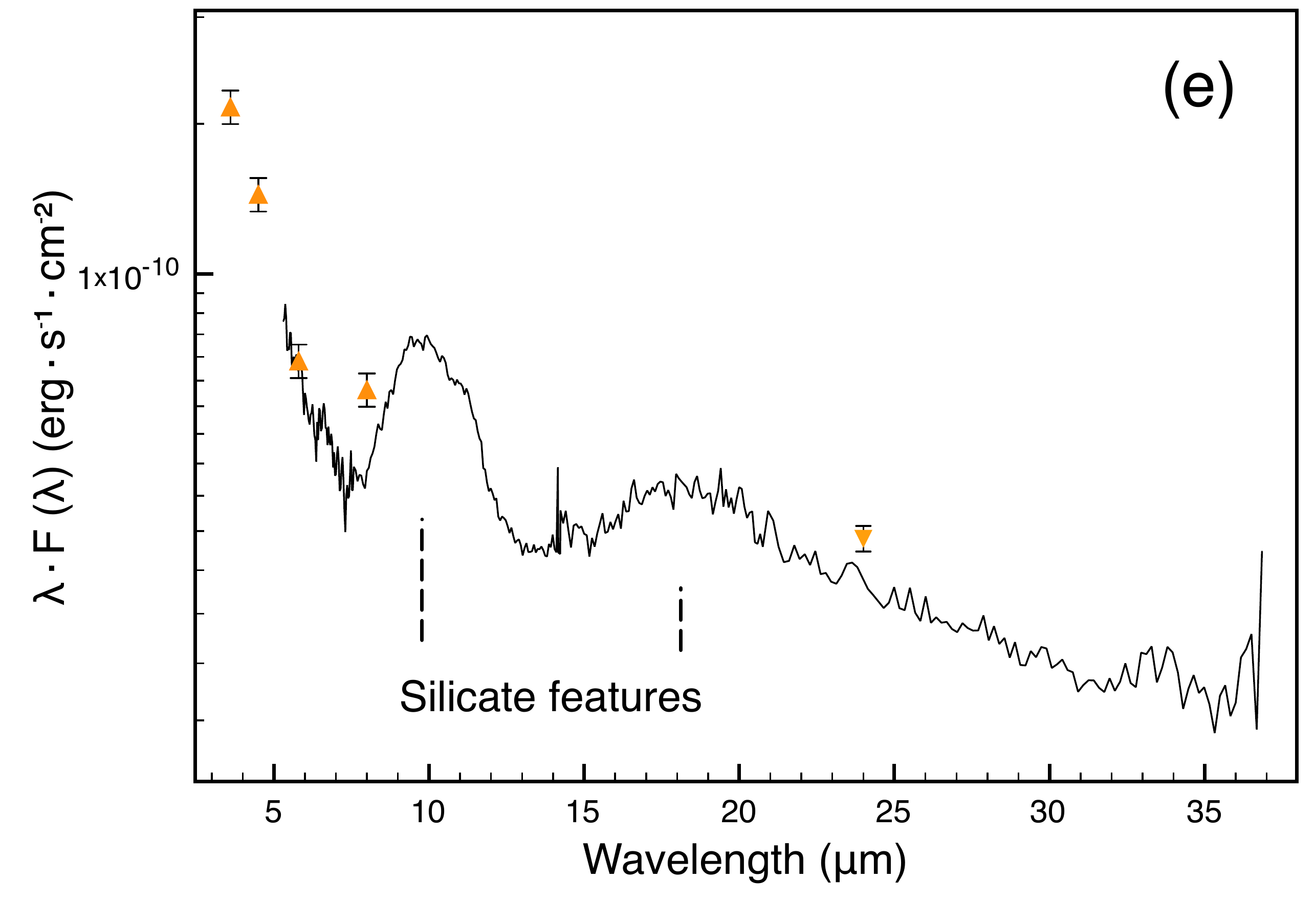}
   \includegraphics[width=7cm]{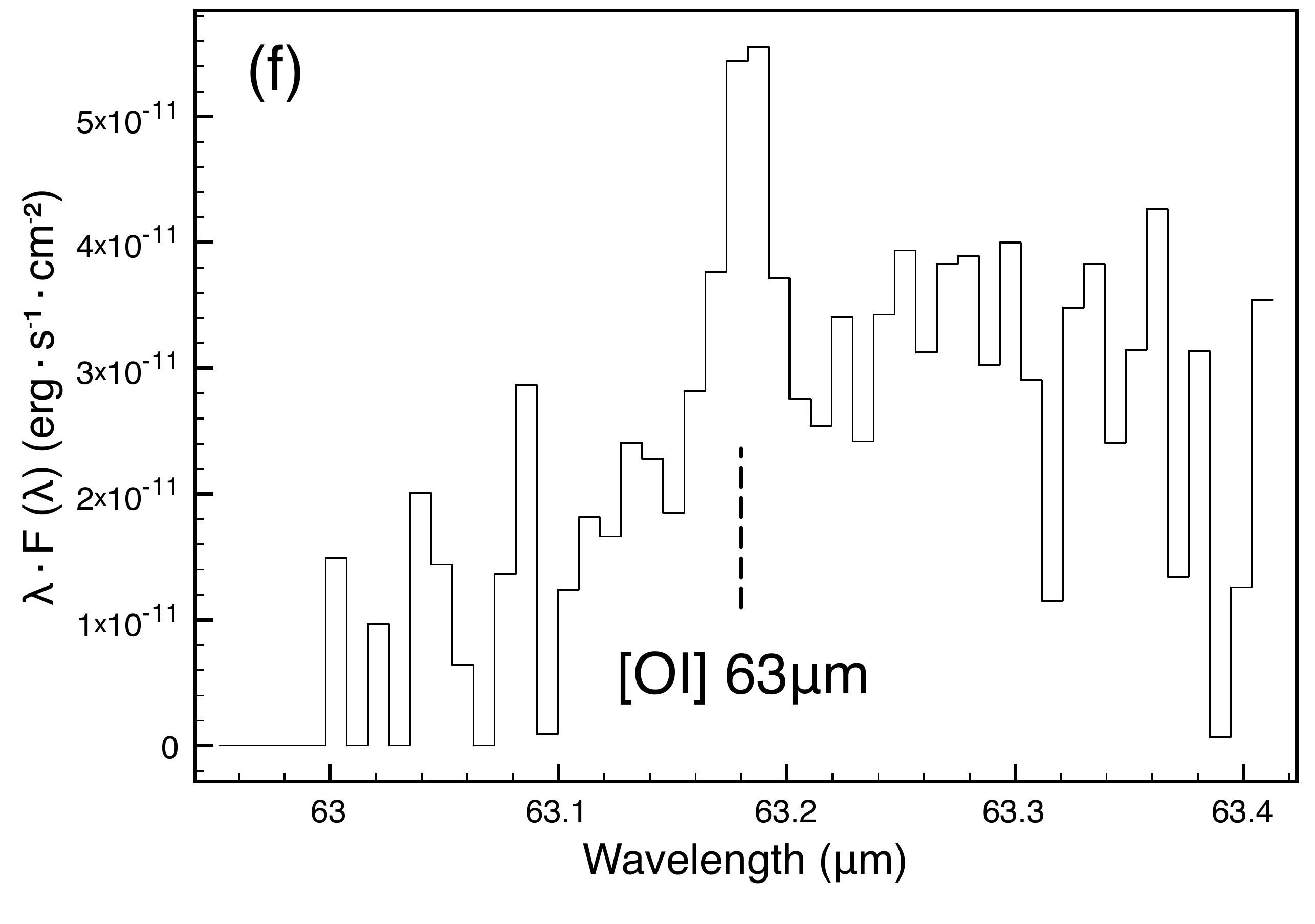}
   \includegraphics[width=13.5cm]{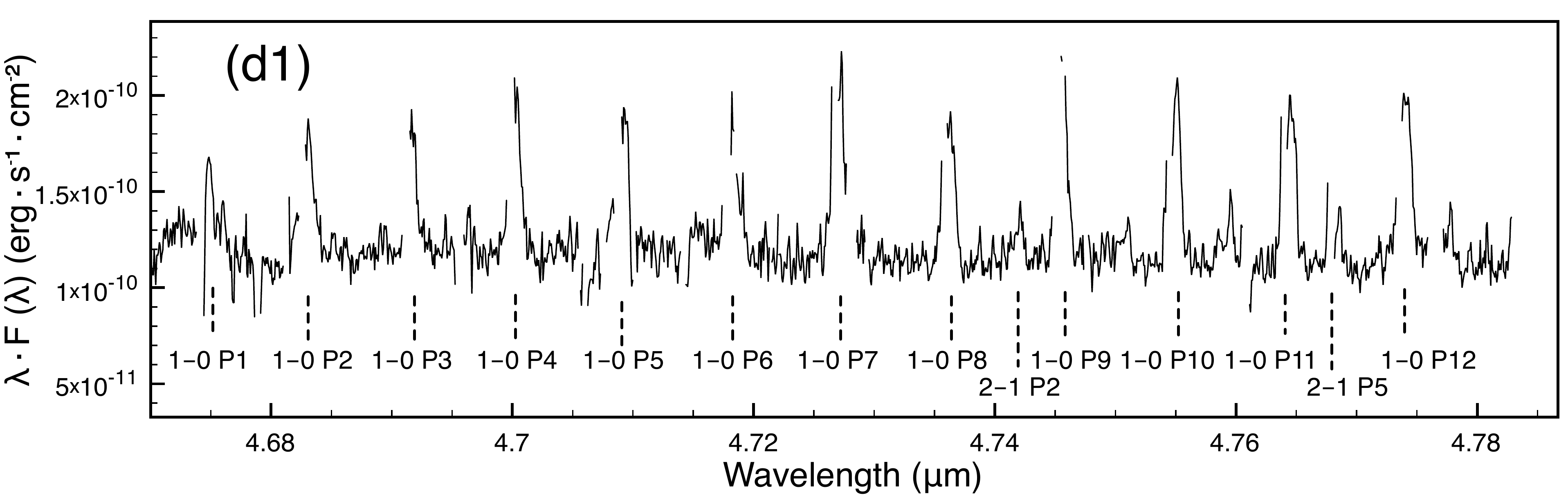}
   \includegraphics[width=13.5cm]{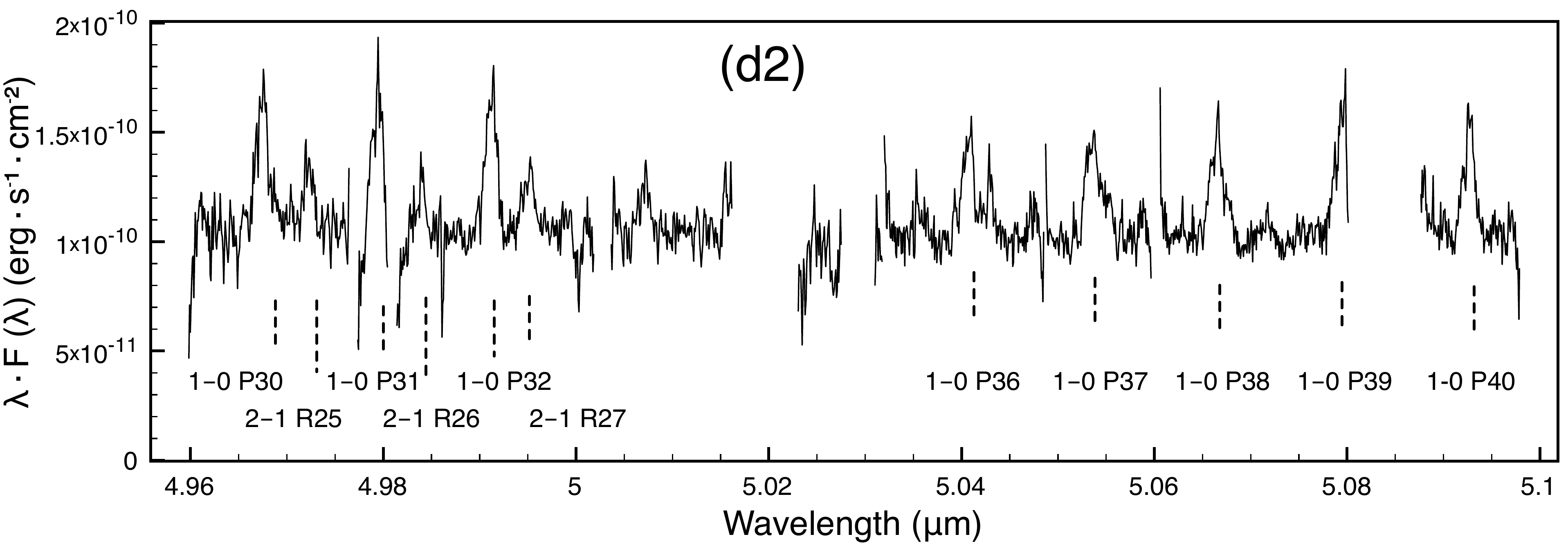}
         \caption{Spectroscopic observations. (a) TNG optical spectrum; (b) WHT J-band spectrum; (c) NOT K-band spectrum; (d) Keck high-resolution spectra; (e) Spitzer MIR spectrum; (f) {\it Herschel} FIR spectrum {of the only detected line ($\rm [O\,{\textsc i}]$ 63 $\mu$m)}; (d1) and (d2) zoom on the CO lines detected in the Keck spectra. Photometric non-simultaneous observations are plotted as color symbols. Spectra shown in (a), (b), (c) were flux-calibrated by means of the photometry.
              }
         \label{Spectra}
   \end{figure*}

Regions in which the transmittance was less than 50\% were masked. The wavelength calibration was determined from the fit to the telluric absorption lines and is generally accurate to within about 0.1 pixels ($\sim$ 0.4 km/s). Flux calibration was performed using the available Spitzer/IRAC {4.5 and 5.8 $\mu$m} photometry (see Table \ref{Photometry} and {Fig.\,\ref{Spectra}d}). In order to obtain the gas velocity with respect to the star, we corrected for the observed radial velocity of FT Tau, as estimated by Guilloteau et al.\ (2013) (${\rm v}_{\rm LSR}=7-9$ km/s).

\subsection{\textit{Herschel}/PACS observations}
FIR spectroscopic observations of FT Tau were obtained with the integral-field spectrometer PACS (Poglitsch et al.\ 2010), on board of the \textit{Herschel} Space Telescope (Pilbratt et al.\ 2010) as part of the \textit{Herschel} Open Time Key Project GASPS (GAS in Protoplanetary Systems, PI: W.\ Dent, {see Dent et al.\ 2013}). {The observations were carried out in the chop- nod mode to remove the background emission and with a single pointing on the source. They cover simultaneously a selected wavelength range in the blue and in the red arms. In particular, the observation acquired in line mode (OBSID: 1342192790) covers the ranges 63.0--63.4 $\mu$m and 180.7--190.3 $\mu$m, with a resolution of 88 {km s$^{-1}$} and 200 km s$^{-1}$. The observation acquired in range mode (OBSID: 1342243501) covers the ranges 71.8--73.3  $\mu$m and 143.5--146.6 $\mu$m, with a spectral resolution of 162 km s$^{-1}$ and 258 km s$^{-1}$.}

{Data were reduced using HIPE 10. Removal of saturated and bad pixels, chop subtraction, flat-field correction, and mean of two nods were performed by means of the available PACS pipeline.}

Photometric FIR observations of FT Tau were also taken with {\it Herschel}/PACS within the GASPS project. The obtained photometric measurements at 70, 100, and 160 $\rm \mu m$ are presented in Howard et al.\ (2013) and shown in Table \ref{Photometry}.  

\subsection{Photometric and Spitzer/IRS observations}
We collected from the literature 44 photometric measurements of FT Tau, from 0.36 $\mu$m to 7 mm (see Table \ref{Photometry} for the fluxes and references). These data have been acquired with 14 different instruments and over more than two decades (see Sect.\,\ref{Variability} and Appendix \ref{Uncertainties} for a further discussion on variability).

A MIR spectrum of FT Tau (spectral range 5.13 $\mu$m - 39.90 $\mu$m) was also retrieved from the literature (Furlan et al.\ 2006).   

 \begin{table*}
      \caption[]{Photometric measurements of FT Tau.}
         \label{Photometry}
     $$ 
         \begin{tabular}{lccccc}
            \hline
            \hline
            \noalign{\smallskip}
            Instrument  & \rm $\lambda$ & Band & Flux & $\lambda F_{\lambda}$ & \\
 & ($\rm \mu m$) & & (publication units) & ($\rm erg \cdot cm^{-2} \cdot s^{-1}$) \\
            \noalign{\smallskip}
            \hline
            \noalign{\smallskip}
            SDSS & 0.36 & u band & 16.681 $\pm$ 0.007 mag $^{\mathrm{\ a}}$  & $(6.69 \pm 0.06) \cdot 10^{-12}$  \\ 
            USNO & 0.44 & B band & 15.48 mag $^{\mathrm{\ b}}$ & $ 1.80 \cdot 10^{-11}$ \\
            SDSS  & 0.48 & g band & 15.804 $\pm$ 0.004 mag $^{\mathrm{\ a}}$   & $ (1.08 \pm 0.01) \cdot 10^{-11}$   \\ 
            USNO & 0.55 & V band & 14.69 mag $^{\mathrm{\ b}}$   &  $ 2.64 \cdot 10^{-11}$ \\
            SDSS &  0.62 & r band & 14.254 $\pm$ 0.004 mag $^{\mathrm{\ a}}$ & $ (3.47 \pm 0.02) \cdot 10^{-11}$ \\ 
            USNO  & 0.68 & R band & 12.06 mag $^{\mathrm{\ b}}$        &  $ 1.07 \cdot 10^{-10}$   \\
            SDSS & 0.76 & i band & 15.023 $\pm$ 0.011 mag $^{\mathrm{\ a}}$ &  $ (6.39 \pm 0.06) \cdot 10^{-11}$  \\ 
            USNO & 0.80 & I band & 11.36 mag $^{\mathrm{\ b}}$ & $ 2.56 \cdot 10^{-10}$ \\
            SDSS  & 0.90 & z band & 12.456 $\pm$ 0.004 mag $^{\mathrm{\ a}}$   &   $ (1.22 \pm 0.01) \cdot 10^{-10}$       \\ 
            2MASS & 1.25 & J band & 10.19 $\pm$ 0.03 mag $^{\mathrm{\ d}}$  & $ (3.15 \pm 0.07) \cdot 10^{-10}$  \\
            2MASS & 1.65 & H band & 9.12 $\pm$ 0.03 mag $^{\mathrm{\ d}}$ & $ (4.58 \pm 0.11) \cdot 10^{-10}$ \\
            2MASS & 2.16 & $\rm K_S$ band & 8.59 $\pm$ 0.02 mag $^{\mathrm{\ d}}$&$ (3.32 \pm 0.06) \cdot 10^{-10}$   \\
            WISE & 3.37 & & 7.75 $\pm$ 0.02 mag $^{\mathrm{\ e}}$          & $ (2.18 \pm 0.04) \cdot 10^{-10}$  \\
            Spitzer/IRAC & 3.6  && 7.64 $\pm$ 0.02 mag $^{\mathrm{\ f}}$    & $ (2.06 \pm 0.04) \cdot 10^{-10}$ \\ 
            Spitzer/IRAC & 3.6  & & 7.89 $\pm$ 0.02 mag $^{\mathrm{\ f}}$ & $ (1.57 \pm 0.07) \cdot 10^{-10}$ \\ 
            Spitzer/IRAC & 4.5  & & 7.12 $\pm$ 0.02 mag $^{\mathrm{\ f}}$ & $ (1.72 \pm 0.03) \cdot 10^{-10}$  \\ 
            Spitzer/IRAC & 4.5  & & 7.44 $\pm$ 0.02 mag $^{\mathrm{\ f}}$  & $ (1.24 \pm 0.06) \cdot 10^{-10}$ \\ 
            WISE & 4.61 & & 7.10 $\pm$ 0.02 mag $^{\mathrm{\ e}}$     &   $ (1.60 \pm 0.03) \cdot 10^{-10}$     \\
            Spitzer/IRAC & 5.8  & & 6.81 $\pm$ 0.02 mag $^{\mathrm{\ f}}$ & $ (1.15 \pm 0.02) \cdot 10^{-10}$ & \\ 
            Spitzer/IRAC & 5.8  & & 7.12 $\pm$ 0.02 mag $^{\mathrm{\ f}}$ & $ (7.91 \pm 0.36) \cdot 10^{-11}$  \\ 
            Spitzer/IRAC & 8.0 & & 5.95 $\pm$ 0.03 mag $^{\mathrm{\ f}}$ & $ (1.00 \pm 0.03) \cdot 10^{-10}$ \\ 
            Spitzer/IRAC & 8.0 & & 6.27 $\pm$ 0.03 mag $^{\mathrm{\ f}}$ & $ (7.32 \pm 0.33) \cdot 10^{-11}$ \\ 
            IRAS & 12 & & 0.46 Jy $\pm$ 14\% $^{\mathrm{\ g}}$    &  $(1.15 \pm 0.16) \cdot 10^{-10}$       \\
            WISE & 12.08 & & 5.09 $\pm$ 0.01 mag $^{\mathrm{\ e}}$  &  $ (7.23 \pm 0.06) \cdot 10^{-11}$         \\
            WISE & 22.19 &&  3.08 $\pm$ 0.02 mag $^{\mathrm{\ e}}$    &  $ (6.59 \pm 0.12) \cdot 10^{-11}$       \\
            Spitzer/MIPS & 24 & & 3.15 $\pm$ 0.04 mag $^{\mathrm{\ f}}$  &  $ (4.90 \pm 0.17) \cdot 10^{-11}$ \\ 
            IRAS &  25 & & 0.65 Jy $\pm$ 12\% $^{\mathrm{\ g}}$ & $ (7.80 \pm 0.94) \cdot 10^{-11}$ \\
            IRAS & 60 & & 0.86 Jy $\pm$ 10\% $^{\mathrm{\ g}}$     &  $ (4.30 \pm 0.43) \cdot 10^{-11}$      \\
	   Spitzer/MIPS & 70  & & 0.28 $\pm$ 0.22 mag $^{\mathrm{\ h}}$  & $ (2.56 \pm 0.47) \cdot 10^{-11}$  \\
             {\it Herschel}/PACS & 70  & & 0.73 $\pm$ 0.07 Jy $^{\mathrm{\ i}}$ & $ (3.08 \pm 0.30) \cdot 10^{-11}$   \\
            IRAS & 100 & & 1.92 Jy $\pm$ 14\% $^{\mathrm{\ g}}$ & $ (5.76 \pm 0.81) \cdot 10^{-11}$ \\
             {\it Herschel}/PACS & 100  & & 0.95 $\pm$ 0.09 Jy $^{\mathrm{\ i}}$ & $ (2.76 \pm 0.27) \cdot 10^{-11}$ \\
             {\it Herschel}/PACS & 160  & & 1.27 $\pm$ 0.19 Jy $^{\mathrm{\ i}}$ & $ (2.34 \pm 0.36) \cdot 10^{-11}$ \\
            CSO  & 350 & & 1106 $\pm$ 82 mJy $^{\mathrm{\ j}}$    &   $ (9.48 \pm 0.70) \cdot 10^{-12}$ \\ 
            JCMT & 450 & & 437 $\pm$ 56 mJy $^{\mathrm{\ j}}$ & $ (2.91 \pm 0.37) \cdot 10^{-12}$  \\ 
            CSO & 624 & & 260 $\pm$ 100 mJy $^{\mathrm{\ k}}$ & $ (1.25 \pm 0.48) \cdot 10^{-12}$ \\ 
            CSO &  769 & & 250 $\pm$ 50 mJy $^{\mathrm{\ k}}$ & $ (9.75 \pm 1.95) \cdot 10^{-13}$ \\ 
            JCMT  & 850 & & 121 $\pm$ 5 mJy $^{\mathrm{\ j}}$   &   $ (4.27 \pm 0.17) \cdot 10^{-13}$  \\ 
            SMA & 880 & & 111 $\pm$ 2 mJy $^{\mathrm{\ l}}$  & $ (3.78 \pm 0.07) \cdot 10^{-13}$ \\ 
            CSO & 1056 & & 137 $\pm$ 40 mJy $^{\mathrm{\ k}}$ & $ (3.89 \pm 1.13) \cdot 10^{-13}$ \\ 
            IRAM  & 1300 & & 130 $\pm$ 14 mJy $^{\mathrm{\ m}}$  & $ (3.00 \pm 0.32) \cdot 10^{-13}$  \\ 
            IRAM & 2700 & & 25 $\pm$ 2.2 mJy $^{\mathrm{\ n}}$  & $ (2.78 \pm 0.24) \cdot 10^{-14}$ \\ 
            VLA & 7000 & & 1.62 $\pm$ 0.27 mJy $^{\mathrm{\ o}}$ & $ (6.94 \pm 1.15) \cdot 10^{-16}$ \\ 
                        \noalign{\smallskip}
            \hline
         \end{tabular}
     $$ 

References: $^{\mathrm{a}}$ Finkbeiner et al.\ 2004; $^{\mathrm{b}}$ Monet et al.\ (2003); $^{\mathrm{d}}$ Cutri et al.\ (2003); $^{\mathrm{e}}$ Wright et al.\ 2010; $^{\mathrm{f}}$ Luhman et al.\ (2010), two observations per wavelength;  $^{\mathrm{g}}$ Beichman et al.\ (1988); $^{\mathrm{h}}$ Rebull et al.\ (2010); $^{\mathrm{i}}$ Howard et al.\ (2013); $^{\mathrm{j}}$ Andrews \& Williams (2005); $^{\mathrm{k}}$ Beckwith \& Sargent (1991); $^{\mathrm{l}}$ Andrews \& Williams (2007); $^{\mathrm{m}}$ Beckwith et al.\ (1990); $^{\mathrm{n}}$ Dutrey et al.\ (1996); $^{\mathrm{o}}$ Rodmann et al.\ (2006).
   \end{table*}


\section{Results from observations} \label{Results}

The spectra obtained by reducing the observations presented in Sect.\,\ref{Observations} are shown in Fig.\,\ref{Spectra}. The optical TNG spectrum (Fig.\,\ref{Spectra}a) shows a number of molecular absorption bands, that are typical of late-type stars, and several emission lines, which are thought to originate from the accretion columns or from the outflow. The J-band WHT spectrum (Fig.\,\ref{Spectra}b) and the K-band NOT spectrum (Fig.\,\ref{Spectra}c) show prominent $\rm Pa\beta$ and $\rm Br\gamma$ emission lines, produced in the accretion process. In the Keck spectra (Fig.\,\ref{Spectra}d, \ref{Spectra}d1, \ref{Spectra}d2) we detected the Pf$\beta$ and Hu$\epsilon$ recombination lines, and CO ro-vibrational lines which are thought to be produced in the disk by thermal excitation or by UV fluorescence. The spectroscopic observations collected with \textit{Herschel}/PACS cover a number of disk tracers, e.g.\ water lines (o-H$_2$O~7$_{07}$-6$_{16}$,  p-H$_2$O~4$_{13}$-3$_{22}$), high-J rotational CO lines (CO J=18-17, and CO J=36-35), the CH$^{+}$ J=5-4, and the $\rm [O\,{\textsc i}]$~$^3$P$_1$-$^3$P$_2$ and $\rm [O\,{\textsc i}]$~$^3$P$_0$-$^3$P$_1$ lines at 63.184 and 145.525 $\mu$m. However, only the $\rm [O\,{\textsc i}]$ 63.184 $\mu$m line was detected in the central spatial pixel (spaxel) of the PACS integral field unit (Fig.\,\ref{Spectra}f), while the other lines remained undetected. For those lines we report the 3$\sigma$ upper limit in Table \ref{Lines}. The Spitzer/IRS spectrum shows prominent silicate features at 10 and 18 $\mu$m (Fig.\,\ref{Spectra}e), which are believed to originate in the optically thin disk surface layer. The properties of all detected lines are listed in Table \ref{Lines}.

\begin{table*}
      \caption[]{Emission lines detected in our spectra and respective vacuum wavelengths. When the line presents instrumental gaps, lower and higher estimates due to the missing region are included in the error. Dereddened fluxes are omitted when the correction for the extinction is negligible.}
         \label{Lines}
     $$ 
         \begin{tabular}{lccccc}
            \hline
            \hline
            \noalign{\smallskip}
            Line & \rm Wavelength& Instrument & Observed flux & Dereddened flux ($A_V=1.8$) & Likely origin \\
             & ($\rm \mu m$) & & ($10^{-14} \ \rm erg \cdot s^{-1} \cdot cm^{-2}$) & ($10^{-14} \ \rm erg \cdot s^{-1} \cdot cm^{-2}$) &  \\
            \noalign{\smallskip}
            \hline
            \noalign{\smallskip}
            H$\beta$ & 0.486 & TNG & $17.9 \pm 0.3$ & 125.5 & Accretion  \\
            Fe \textsc{ii} & 0.492 & TNG & $0.73 \pm 0.03$ & 4.9 & Accretion \\
            Fe \textsc{ii}  & 0.502 & TNG & $0.83 \pm 0.03$ & 5.6 & Accretion \\
            Fe \textsc{ii} & 0.517 & TNG & $1.65 \pm 0.07$ & 7.1 & Accretion \\
            He \textsc{i} & 0.588 & TNG & $2.63 \pm 0.31$ & 12.1 & Accretion \\
            NaD &  0.589 & TNG & $0.62 \pm 0.03$ & 4.8 & Accretion \\
            NaD  & 0.590 & TNG & $0.34 \pm 0.03$ & 2.7 & Accretion \\
            $\rm [O\,{\textsc i}]$ & 0.630 & TNG & $0.68 \pm 0.21$ & 2.6 & Disk / Outflow \\
            H$\alpha$ & 0.654 & TNG & $169.4 \pm 0.6$ & 679.8 & Accretion \\
            He \textsc{i} & 0.668 & TNG & $1.69 \pm 0.05$ & 6.9 & Accretion \\
            Pa$\beta$ & 1.282 & WHT & $30.5 \pm 0.7$  & 46.0 & Accretion \\
            Br$\gamma$ & 2.166 & NOT & $12.5 \pm 0.8$  & 14.6 & Accretion \\
            Pf$\beta$ & 4.654 & Keck & $4.45 \pm 0.21$  & - & Accretion  \\
            CO 1$-$0 P1 & 4.674 & Keck & $0.30 \pm 0.11$  & - & Disk  \\
            \vspace{1mm}
            CO 1$-$0 P2 & 4.683 & Keck & $1.16 ^{+0.13} _{-0.28} $   & - & Disk     \\
            \vspace{1mm}
            CO 1$-$0 P3 & 4.691 & Keck & $1.39 ^{+0.17} _{-0.51} $  & - &  Disk \\
            \vspace{1mm}
            CO 1$-$0 P4 & 4.699 & Keck & $1.65 ^{+0.18} _{-0.38} $ & - & Disk  \\
            \vspace{1mm}
            CO 1$-$0 P5 & 4.709 & Keck & $1.25 ^{+0.23} _{-0.28} $ & - & Disk  \\
            \vspace{1mm}
            CO 1$-$0 P6 & 4.718 & Keck & $1.48 ^{+0.17} _{-0.22} $  & - & Disk \\
            \vspace{1mm}
            CO 1$-$0 P7 & 4.727 & Keck & $1.97 ^{+0.14} _{-0.18} $ & - & Disk \\
            \vspace{1mm}
            CO 1$-$0 P8 & 4.736 & Keck & $0.90 ^{+0.08} _{-0.10} $  & - & Disk   \\
            CO 2$-$1 P2 & 4.741 & Keck & $0.26 \pm 0.13$ & - & Disk   \\
            CO 1$-$0 P9 & 4.745 & Keck & $1.50 ^{+0.51} _{-0.71} $ & - & Disk  \\
            CO 1$-$0 P10 & 4.756 & Keck & $1.58 \pm 0.12$  & - & Disk   \\
            CO 2$-$1 P4 & 4.759 & Keck & $0.28 \pm 0.10$  & - & Disk  \\
            CO 1$-$0 P11 & 4.764 & Keck & $0.66 \pm 0.13$ & - & Disk  \\
            CO 2$-$1 P5 & 4.768 & Keck & $0.49 \pm 0.07$  & - & Disk    \\
	   CO 1$-$0 P12 & 4.774 & Keck & $1.62 ^{+0.12} _{-0.24} $ & - & Disk   \\
	   CO 1$-$0 P30 & 4.967 & Keck & $1.01 \pm 0.19$ & - & Disk   \\
            CO 2$-$1 R25 & 4.971 & Keck & $0.41 \pm 0.14$  & - & Disk  \\
            CO 1$-$0 P31 & 4.979 & Keck & $1.01 \pm 0.26$ & - & Disk \\
            CO 1$-$0 P32 & 4.991 & Keck & $0.91 \pm 0.21$ & - & Disk \\
            CO 2$-$1 R27 & 4.995 & Keck & $0.25 \pm 0.10$ & - & Disk \\
            CO 1$-$0 P36  & 5.041 & Keck & $0.83 \pm 0.08$   & - & Disk   \\
            CO 2$-$1 P30 & 5.043 & Keck & $0.20 \pm 0.07$  & - & Disk \\
            CO 1$-$0 P37 & 5.053 & Keck & $0.61 \pm 0.13$ & - & Disk    \\
            CO 1$-$0 P38 & 5.066 & Keck & $0.71 \pm 0.14$ & - & Disk  \\
            CO 1$-$0 P39 & 5.079 & Keck & $0.88 \pm 0.18$    & - & Disk      \\
            CO 1$-$0 P40 & 5.092 & Keck & $0.86 \pm 0.09$  & - & Disk  \\
            $\rm [O\,{\textsc i}]$ & 63.185 & {\it Herschel} & $1.6 \pm 0.5$ & - & Disk / Outflow  \\
            $\rm o-H_2O$ & 71.947 & {\it Herschel} & $<1.41$ & - & - \\
            CH+ $5-4$ & 72.140 & {\it Herschel} & $<1.37$ & - & - \\
            CO $36-35$ & 72.843 & {\it Herschel} & $<0.97$ & - & - \\
            $\rm p-H_2O$ & 144.518 & {\it Herschel} & $<0.32$ & - & - \\
            CO $18-17$ & 144.784 & {\it Herschel} & $<0.33$ & - & - \\
            $\rm [O\,{\textsc i}]$ & 145.525 & {\it Herschel} & $<0.31$ & - & -  \\
                           \noalign{\smallskip}
            \hline
         \end{tabular}
     $$ 

   \end{table*}

In this section we describe the methods applied to derive the stellar properties (Sect.\,\ref{Results_stellar}), a few disk properties (Sect.\,\ref{Results_disk}), and the mass accretion and outflow rates (Sect.\,\ref{Results_accretionoutflow}).

\subsection{Stellar properties} \label{Results_stellar}
First, we determined the spectral type of FT Tau by comparing the optical TNG spectrum with the spectra of the MILES stellar libraries (Sanchez-Blazquez et al.\ 2006). The observed absorption features suggest that FT Tau is an M2 or M3 star, in agreement with the result by Rebull et al.\ (2010). {The temperature scale is based on Cohen \& Kuhi (1979).} Then, we adopted a PHOENIX model ($T_{\rm eff}=3400 \ {\rm K}$, $\log(g)=3.5$, $\rm [Fe/H]=0.0$) of the stellar atmosphere (Hauschildt et al.\ 1999) to reproduce the stellar spectrum of the source. 

The optical and NIR stellar spectra may suffer from strong extinction {by the dust along the line of sight, either foreground or in the disk}. We estimated the visual extinction, $A_V$, by comparing the colors of the adopted PHOENIX model spectrum with the available photometry. Since the UV excess of young accreting stars may extend up to red optical wavelengths and the IR excess may start at $\sim \rm 2 \ \mu m$, we used (J-H) colors. Using the extinction law by Cardelli et al.\ (1989) and $R_V=3.1$, we obtained an extinction of $A_V=1.8$. For a discussion on the uncertainties affecting the determination of the extinction, see Appendix \ref{Uncertainties}.

To determine the stellar luminosity and radius we imposed the reddened PHOENIX flux at 1.25 $\rm \mu m$ to be equal to the observed one, as available from 2MASS (see Table \ref{Photometry}), assuming that the latter is only due to photospheric emission. 

The estimated radius is $R_*=1.7 \ {\rm R_{\odot}}$ and the luminosity $L_*=0.35 \ {\rm L_{\odot}}$. Uncertainties on those estimates are due to the distance ($140 \pm 10$ pc, Kenyon et al.\ 1994) and effective temperature {(3\% in the M2-M4 range, Kenyon \& Hartmann 1995)}. 

Finally, using pre-main sequence (PMS) evolutionary tracks by Siess et al.\ (2000) and assuming [Fe/H]=0.0, we estimated the mass and the age of the source ($M_* = 0.3 \ {\rm M}_{\odot}$ and age $1.6 \cdot 10^6$ yr). The derived stellar properties (see Table \ref{Est_properties}) are typical for TTSs (Beckwith et al.\ 1990, Kenyon \& Hartmann 1995, Hartigan et al.\ 1995).

\begin{table}
      \caption[]{Estimated stellar properties.}
         \label{Est_properties}
     $$ 
         \begin{tabular}{lcccc}
            \hline
            \hline
            \noalign{\smallskip}
            Visual extinction & $A_V$ & 1.8 $\pm$ 0.6 \\
            Spectral type & & M3 $\pm$ 1 \\
            Effective temperature & $T_{\rm eff}$ & 3400 $\pm$ 200 K \\
            Luminosity & $L_*$ & $0.35 \pm 0.09$ \ L$_{\odot}$ \\
            Radius & $R_*$ & $1.7 \pm 0.2$ \ R$_{\odot}$ \\
            Mass & $M_* $ & $0.3 \pm 0.1$ \ M$_{\odot}$ \\
            Age & & $1.6 \pm 0.3$ \ Myr \\
            \noalign{\smallskip}
            \hline
            \end{tabular}
     $$ 

   \end{table}

\subsection{Disk properties} \label{Results_disk}

\begin{figure*}
   \centering
   \includegraphics[width=14cm]{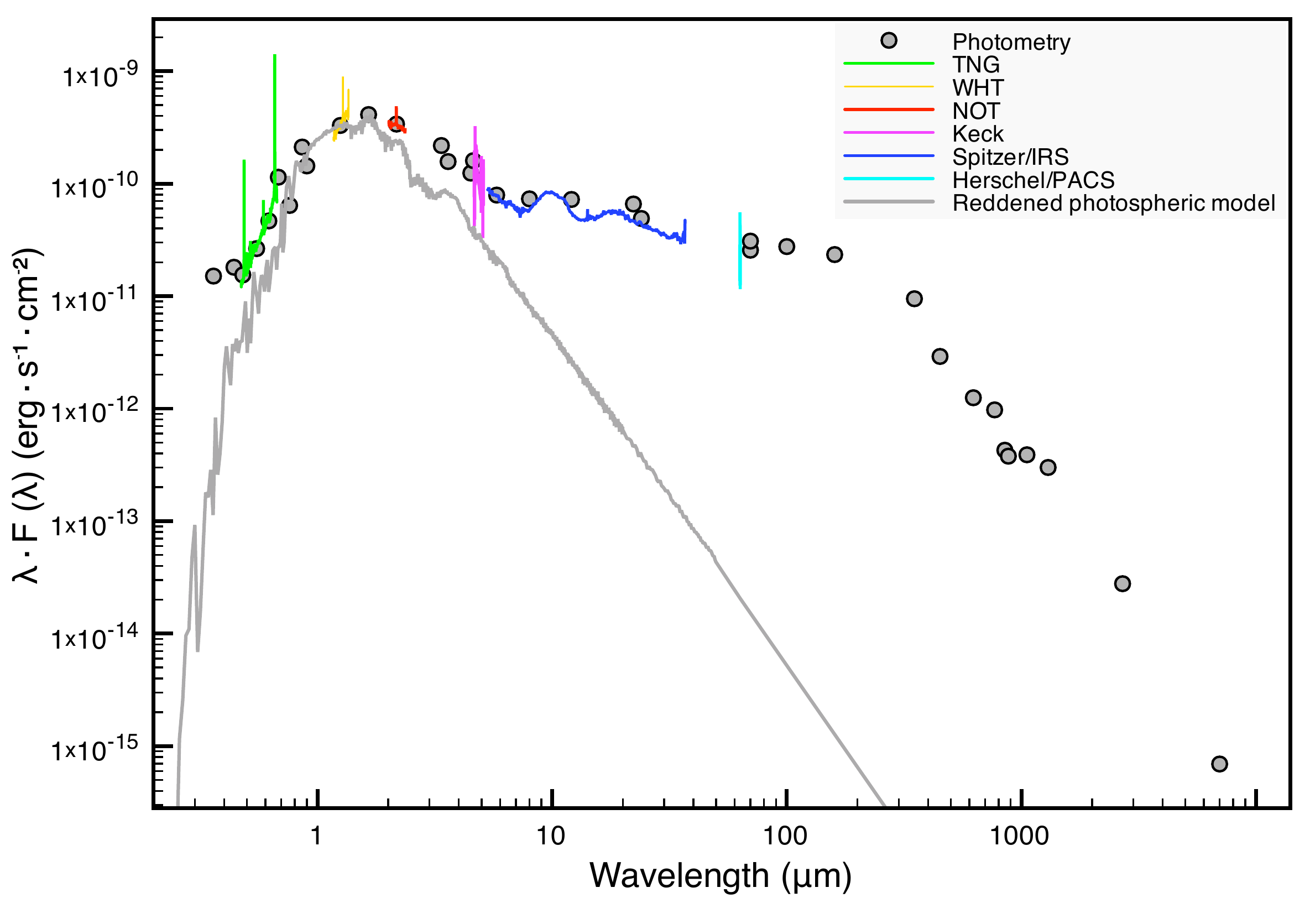}
     \caption{SED of FT Tau. All the available photometric and spectroscopic non-simultaneous observations are plotted. The reddened \textit{Phoenix} model is reproducing the stellar photospheric emission (grey line). The SED clearly shows UV/optical excess emission at wavelengths shorter than $\sim$ 0.8 $\rm \mu m$ and infrared excess beyond $\sim$ 2 $\rm \mu m$. IRAS photometry is excluded and the higher IRAC 5.8 and 8 $\mu$m photometry is omitted giving more weight to the IRS spectrum. Error bars are not visible at this scale.  
              }
         \label{SED}
   \end{figure*}

\subsubsection{Disk flaring and {dust composition}} \label{Analysis_flaring}
FT Tau shows a very prominent IR excess (see Fig.\ \ref{SED}). By integrating the excess emission over the photosphere, we obtained an IR excess luminosity $L_{\rm IR} = 0.12 \pm 0.01 \ {\rm L}_{\odot}$. 

We also measured the IR excess at different wavelengths (see Table \ref{Excess}) by subtracting the photospheric model from the observed photometry. This provides qualitative information on the geometry of the disk. As the dust grows and settles toward the mid-plane, the vertical scale height of the disk decreases, causing less reprocessing of the stellar radiation and, thus, a smaller MIR excess. According to the evolutionary scheme by Fang et al.\ (2009), the excess shown by FT Tau is typical of objects with disks that are evolving from a \textit{mildly flaring} to a \textit{flat} geometry.

{The MIR spectrum of FT Tau (see Fig.\,\ref{Spectra}e) shows prominent, narrow and smooth silicate features. These are thought to originate in the warm, optically thin disk surface and provide information on the silicate dust in this layer. As shown by Bouwman et al.\ (2001) for Herbig Ae/Be stars, silicate features peaking at $\sim$ 10 $\mu$m, as in the case of FT Tau, are indicative of a dust population dominated by grains as small as 0.1 $\mu$m. The flattening of these features (see Furlan et al.\ 2006 for a large sample of TTSs) can be a tracer of the evolution of the dust population at the disk surface. Some processes such as stellar winds and radiation pressure can deplete sub-$\mu$m size grains (Olofsson et al.\ 2009). The narrow and prominent nature of the silicate features shown by FT Tau suggests that these processes are not yet efficient in this disk. The 10 $\mu$m feature can also provide insight into the crystallinity of the silicate (e.g.\ Sargent et al.\ 2006). The absence of substructure in the MIR spectrum of FT Tau indicates that the silicates are mostly amorphous.}

Finally, the MIR spectrum does not show Polycyclic Aromatic Hydrocarbon (PAH) emission features. PAH emission is indeed hardly detected in TTSs (Furlan et al.\ 2006), while it is common in more massive Herbig Ae/Be stars (see e.g.\ Meeus et al.\ 2001). This can be explained either in terms of different grain composition or the weaker UV radiation field of low-mass stars. 

\begin{table}
      \caption[]{Infrared excess with respect to the photospheric model measured at different wavelengths. Reported errors are due to the instrumental errors and to different measurements from different observations, where available.}
         \label{Excess}
     $$ 
         \begin{tabular}{lccc}
            \hline
            \hline
            \noalign{\smallskip}
            Wavelength ($\rm \mu m$) & Excess (mag) \\
            \hline
            \noalign{\smallskip}
            3.6 & 0.83 $\pm$ 0.12 \\
            4.5 & 1.37 $\pm$ 0.16 \\
            5.8 & 1.49 $\pm$ 0.16 \\
            8.0 & 2.56 $\pm$ 0.16 \\
            24.0 & 5.26 $\pm$ 0.04 \\
            \noalign{\smallskip}
            \hline
            \end{tabular}
     $$ 

   \end{table}

\subsubsection{Disk inner radius} \label{Results_diskCO}
Most of the CO ro-vibrational lines detected with Keck/NIRSPEC are from transitions from the first vibrational level ($\nu = 1 - 0$) and their fluxes are typically a factor of a few higher than those from $\nu = 2 - 1$ (see Table \ref{Lines}).

\begin{figure}
   \includegraphics[width=8cm]{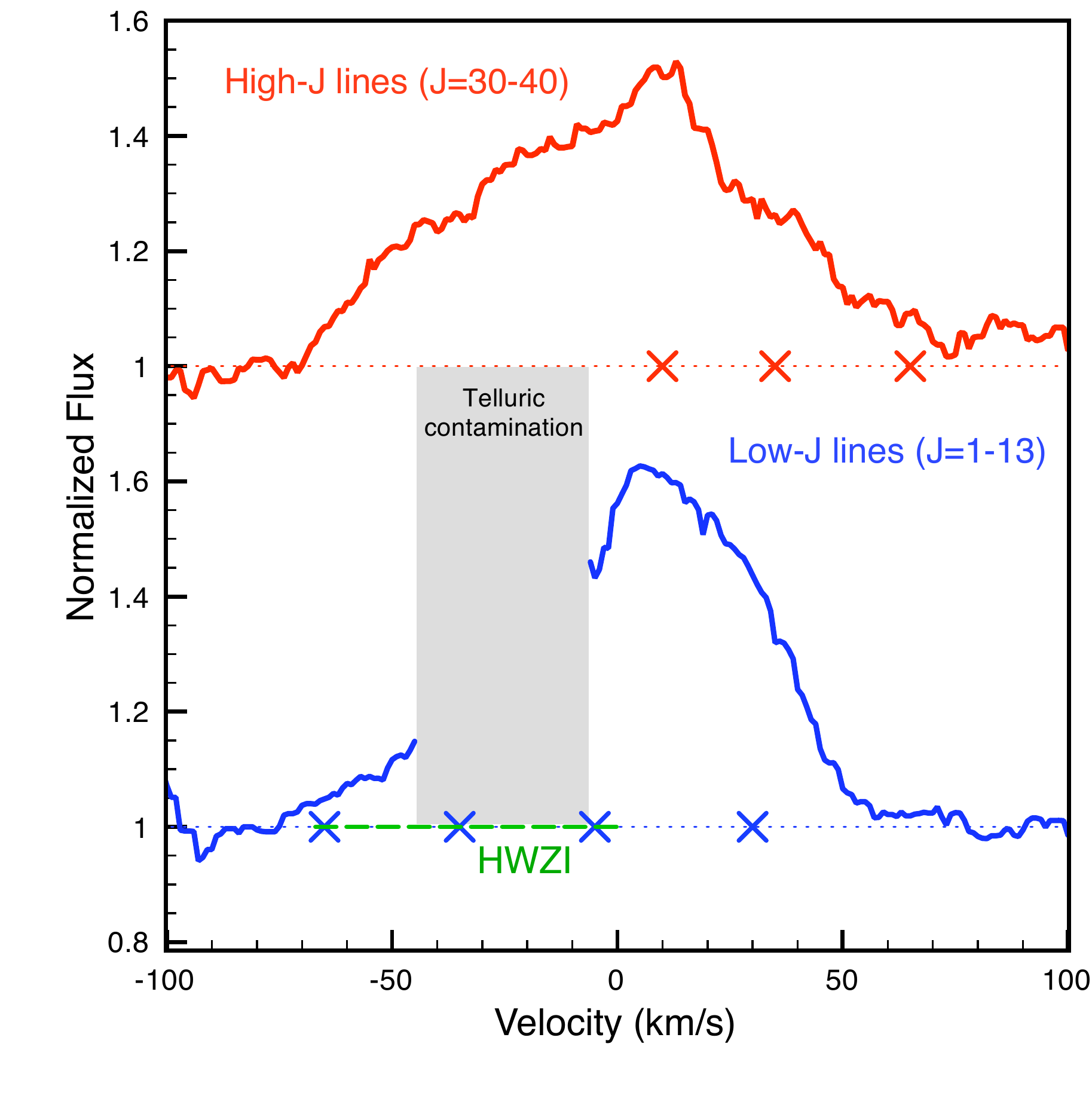}
     \caption{Average profiles of the CO $\nu=1-0$ high-J (red line) and low-J (blue line) ro-vibrational lines. The green dashed line indicates the $\rm (HWZI)_{obs}$ of the lines. The red and blue crosses indicate the position of $\nu=2-1$ blended lines.}
         \label{CO_sum}
   \end{figure}

All $\nu = 1 - 0$ low-J (up to J=12) lines are strongly contaminated by atmospheric absorption/emission lines which does not allow to recover the full line profile. The uncertainty on the line fluxes is obtained by assuming a lower/upper flux limit equal to the line intensity at the edges of the region affected by the telluric lines. On the contrary, $\nu = 1 - 0$ high-J (from J=30 to 40) lines do not suffer from telluric contamination, and their profiles are strongly asymmetric toward the red (see Fig.\ \ref{CO_sum}). A likely explanation for this asymmetry is that most of these lines are blended with $\nu = 2 - 1$ lines. 

The disk inner radius can be estimated from the width of the CO line profile after summing over all the detected lines and correcting for the instrumental profile:
\begin{equation} \label{Eq_innerradius}
R_{\rm in} = \frac{GM_*}{(\Delta V_{\rm obs} / \sin {\textit i})^2} \simeq 0.065 \cdot (\sin {\textit i})^2 \ {\rm AU}
\end{equation}
where $M_*$ is the stellar mass, $\Delta V_{\rm obs} = 65$ km/s is the Half Width at Zero Intensity (HWZI) of the CO profile, and $i$ is the disk inclination (see Sect.\,\ref{Modeling_inclination}). 
 
\subsection{Mass accretion and mass outflow rate} \label{Results_accretionoutflow}

\subsubsection{Mass accretion rate} \label{Results_accretion}
As shown by e.g.\ Pringle (1981), the accretion luminosity, $L_{\rm acc}$, released in accretion disks or boundary layers, is related to the mass accretion rate, $\dot{M}_{\rm acc}$, and depends on the assumed width over which the emission occurs (see e.g.\ Bertout, Basri, \& Bouvier 1989). In this work, we consider the accretion luminosity released in the impact of the accretion flow, as:
\begin{equation} 
L_{\rm acc} \simeq \left( 1-\frac{R_*}{R_{\rm in}} \right) \frac{GM_*}{R_*}\dot{M}_{\rm acc}
\end{equation}
(Gullbring et al.\ 1998) where $R_*$ and $M_*$ are the stellar radius and mass, and $R_{\rm in}$ is the disk inner radius. We were not able to directly measure the accretion luminosity since the UV region is only partially covered by the available observations. Therefore, we estimated the mass accretion rate by employing observed empirical correlations between $L_{\rm acc}$ and the luminosity of optical and NIR emission lines which are thought to be excited in the accretion columns (Fang et al.\ 2009, Muzerolle et al.\ 1998c) such as H$\alpha$, H$\beta$, He \textsc{i}, Br$\gamma$, and Pa$\beta$ (see Table \ref{Lines}). Furthermore, we estimated the accretion luminosity from the visual excess with respect to the stellar photosphere as in Hartigan et al.\ (1995).

The derived estimates are in agreement within a factor 2, with an average value of $L_{\rm acc}=0.15 \ {\rm L}_{\odot}$, corresponding to $\dot{M}_{\rm acc}=3.1 \cdot 10^{-8} \ {\rm M}_{\odot}/{\rm yr}$ (see Table \ref{Accretion}). The largest uncertainties on the estimated $L_{\rm acc}$ and $\dot{M}_{\rm acc}$ are due to the scattering of the empirical correlations. On the contrary, the errors on $L_{\rm acc}$ obtained from the visual excess are due to the continuum determination and hence to the uncertainty on the estimated $A_V$. Further uncertainty may be due to the employed bolometric corrections. 

\begin{table}
      \caption[]{Accretion luminosity and mass accretion rate estimates from optical/NIR line luminosities and from optical excess. {References: (1) Fang et al.\ (2009); (2) Muzerolle et al.\ (1998c); (3) Eq.\ (3) of Hartigan et al.\ (1995).}}
         \label{Accretion}
     $$ 
         \begin{tabular}{lcccc}
            \hline
            \hline            
            \noalign{\smallskip}
            Method & $L_{\rm acc}$ & $\dot{M}_{\rm acc}$ & Ref. \\
             & ($\rm L_{\odot}$) & ($\rm 10^{-8} \ M_{\odot}/yr$) & \\ 
            \hline
            \noalign{\smallskip}
            \vspace{1mm}
            H$\beta$ luminosity & $0.09 \ ^{+0.11}_{-0.05}$ & $1.9 \ ^{+2.4}_{-1.1}$ & (1) \\
            \vspace{1mm}
            He \textsc{i} luminosity & $0.19 \ ^{+0.78}_{-0.15}$ & $4.1 \ ^{+17}_{-3.2}$ & (1) \\
            \vspace{1mm}
            H$\alpha$ luminosity & 0.17 $^{+0.26}_{-0.10}$ & $3.7 \ ^{+5.7}_{-2.2}$ & (1) \\
            \vspace{1mm}
            Pa$\beta$ luminosity & 0.11 $^{+0.31}_{-0.08}$ & $2.4 \ ^{+6.8}_{-1.7}$ & (2) \\
            \vspace{1mm}
            Br$\gamma$ luminosity & 0.19 $^{+1.11}_{-0.16}$ & $4.1 \ ^{+24}_{-3.4}$ & (2) \\
            \vspace{1mm}
            Visual excess & 0.12 $^{+0.03}_{-0.02}$ & $2.6 \ ^{+0.6}_{-0.5}$ & (3) \\
            \noalign{\smallskip}
            \hline
            \end{tabular}
     $$ 

   \end{table}

\subsubsection{Mass outflow rate}
Optical and IR forbidden lines (e.g.\ atomic oxygen lines) are typical jet tracers. Following the correlation found by Hollenbach (1985), the $\rm [O\,{\textsc i}] \ 63 \ \mu m$ line is commonly used to constrain the mass outflow rate (see e.g.\ Ceccarelli et al.\ 1997, Podio et al.\ 2012). A similar correlation has been found for the optical $\rm [O\,{\textsc i}] \ 6300 \ \AA$ (see e.g.\ Hartigan et al.\ 1995).

The $\rm [O\,{\textsc i}] \ 63 \ \mu m$ from FT Tau was detected only in the central spaxel (see Sect.\,\ref{Results}). Thus, the line originates in a region around the source smaller than $\sim$ 1,300 AU and we cannot exclude a priori that a significant fraction of it originates in the disk.

Howard et al.\ (2013) found a tight correlation between the flux of the $\rm [O\,{\textsc i}] \ 63 \ \mu m$ and the continuum flux at $\rm 63 \ \mu m$ for Taurus sources showing no evidence of outflow. Jet sources instead show a line flux exceeding the value predicted by the correlation by up to two orders of magnitude, indicating that a significant fraction of the emission is produced in the jet/outflow. 

{The correlation by Howard et al. (2013) indicates that for FT Tau, up to 85\% of the observed $\rm [O\,{\textsc i}] \ 63 \ \mu m$ line flux could originate in the disk. Similarly, also a fraction of the observed $\rm [O\,{\textsc i}] \ 6300 \ \AA$ flux could be produced in the disk. Thus, we used the $\rm [O\,{\textsc i}] \ 6300 \ \AA$ and the $\rm [O\,{\textsc i}] \ 63 \ \mu m$ line luminosity and the correlation by Hollenbach (1985) and Hartigan et al.\ (1995) to derive an upper limit on the mass outflow rate $\dot{M}_{\rm W}$ (see Table \ref{Outflow}). We found that $\dot{M}_{\rm W}<8.9 \cdot 10^{-10} \ {\rm M}_{\odot}/{\rm yr}$. Another uncertainty of the $\rm [O\,{\textsc i}] \ 63 \ \mu m$ flux can be source variability, since our optical spectrum has been flux-calibrated through photometry (see Sect.\,\ref{Variability}).}

\begin{table}
      \caption[]{Mass outflow rate estimate. {References: (1) Eq.\ (A11) of Hartigan et al.\ (1995); (2) Eq.\ (A13) of Hartigan et al.\ (1995)}}
         \label{Outflow}
     $$ 
         \begin{tabular}{lcccc}
            \hline
            \hline
            \noalign{\smallskip}
            Method & $\dot{M}_{\rm W}$ & Ref.\\
             & ($\rm 10^{-10} \ M_{\odot}/yr$) & \\
            \hline
            \noalign{\smallskip}
            \vspace{1mm}
            $\rm [O\,{\textsc i}] \ 6300 \ \AA$ luminosity & $< 7.2$ & (1) \\
            $\rm [O\,{\textsc i}] \ 63 \ \mu m$ luminosity & $< 10.7$ & (2) \\
            \noalign{\smallskip}
            \hline
            \end{tabular}
     $$ 

   \end{table}
   
\section{Modeling the disk of FT Tau} \label{Modeling}
To interpret the SED and the available line emission from the disk, we use the Monte Carlo radiative transfer code MCFOST (Pinte et al.\ 2006) and the thermo-chemical disk modeling code \textit{ProDiMo} (Woitke et al.\ 2009, Kamp et al.\ 2010) sequentially. We fix the stellar properties as estimated from the data analysis (see Sect.\,\ref{Results_stellar}) and run MCFOST to determine a set of dust properties that reproduces the observed SED. We perform a $\chi ^2$ minimization of the SED and obtain a base set of parameters. As a second step, we run a grid of \textit{ProDiMo} models to study the behavior of predicted line fluxes by comparing the results with the observations. The results of the dust and gas modeling are discussed in Sect.\,\ref{Modeling_Mcfost} and Sect.\,\ref{Modeling_Prodimo} respectively.   

\subsection{Dust modeling with MCFOST} \label{Modeling_Mcfost}
MCFOST calculates thermal and chemical properties of the dust disk by treating grains as spherical and homogeneous particles. It is based on the Monte Carlo method, allowing monochromatic photon packets to propagate through the circumstellar environment. Both photospheric emission and dust thermal emission are considered as radiation sources. 

{As shown in Sect.\,\ref{Results_accretion} and Table \ref{Accretion}, observed emission line luminosities suggest $L_{\rm acc}$ values between 0.09 and 0.19 $\rm L_{\odot}$. To study the impact of the UV luminosity, we assumed in the models an average value of $L_{\rm acc}=0.15 \ {\rm L}_{\odot}$. Then, to explore possible variability we also assumed a value three times lower. To translate from accretion luminosity to $f_{\rm UV}=L_{\rm UV}(90-250 \ {\rm nm})/L_*$, we assumed a blackbody spectrum at 10,000 K for the accretion shock. This yields $f_{\rm UV}=0.07$ and 0.025 (in the following denoted as high and low $f_{\rm UV}$). In the models, the UV spectrum in the narrow range between 90 and 250 nm is approximated by a power-law $F_{\lambda} \propto \lambda^{p_{\rm UV}}$ with ${p_{\rm UV}}=2.0$.} 

\begin{figure}
   \includegraphics[width=9cm]{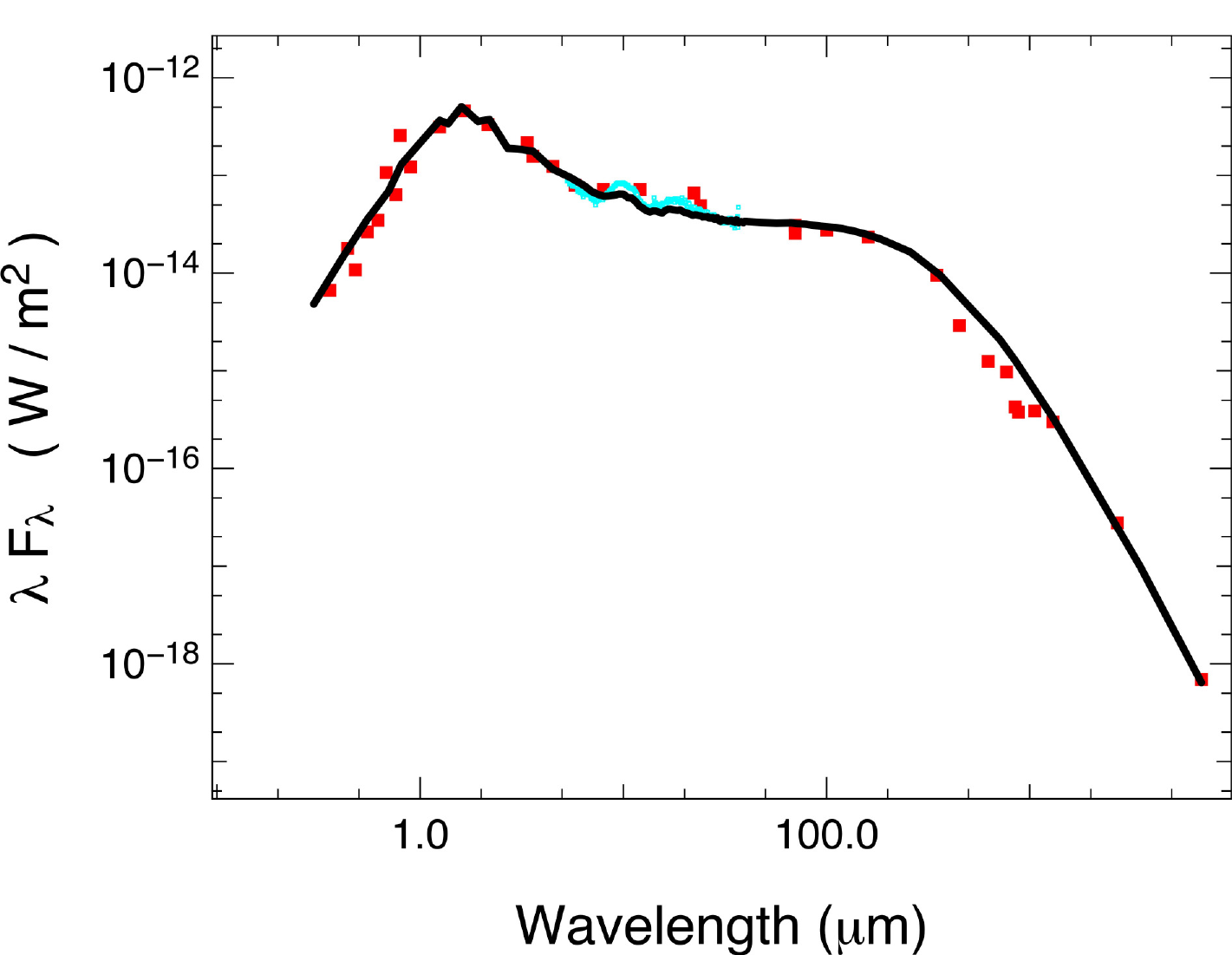}
     \caption{SED predicted by MCFOST using the input parameters listed in Table \ref{Model} (black solid line) overplotted on the observed photometric points (red dots). The blue spectrum is the Spitzer/IRS spectrum.}
         \label{Mcfost_sed}
   \end{figure}

It is well known that pure SED modeling is highly degenerate and thus we choose a strategy where we fix a couple of input parameters to reasonable values, {perform a crude exploration of disk scale height, dust composition, settling and inclination, and perform a $\chi ^2$ minimization leaving only the disk dust mass and optical extinction as} free parameters. The fixed/explored parameters and the results from the fitting are listed in Table \ref{Model} (see Fig.\ \ref{Mcfost_sed}). The $\chi ^2$ fit of the SED results in an optical extinction $A_V=1.6$ in agreement with the result inferred by photometric colors. We obtain a disk dust mass $M_{\rm d}=\rm 9 \cdot 10^{-4} \ M_{\odot}$. {The outer radius is poorly constrained and we explore in the following the impact of three different outer radii, 50, 100 and 200 AU} (see Sect.\,\ref{Oxygen_line} for a discussion). Below, we discuss some modeling aspects and degeneracies in more detail.

\begin{table*}
      \caption[]{Parameters of the FT Tau model. The difference between an `explored' and `free' parameter is that the former is set after an exploratory parameter study while the latter is derived using $\chi^2$ fitting of the SED.}
         \label{Model}
     $$ 
         \begin{tabular}{lccc}
            \hline
            \hline
            Parameter & Symbol & Comments & Value  \\
            \hline
            \noalign{\smallskip}
            Stellar luminosity & $L_{*}$ & Derived from observations & 0.35 $\rm L_{\odot}$ \\ %
            Stellar mass & $M_{*}$ & '' & 0.3 $\rm M_{\odot}$ \\ %
            Stellar radius & $R_{*}$ & '' & 1.7 $\rm R_{\odot}$ \\ %
            Effective temperature & $T_{\rm eff}$& '' & 3400 K \\ %
            Distance & $d$ & '' & 140 pc \\ %
            Slope of UV excess distribution & $p_{\rm UV}$ & Fixed in MCFOST & 2.0 \\ %
            Slope of grain size distribution & $a_{\rm pow}$ & '' & 3.5 \\ %
            Dust mass density & $\rm \rho_{d}$ & '' & 3.5 $\rm g \ cm^{-3}$\\ %
            {Slope of surface mass density} & $\epsilon$ & '' & $-1$ \\
            Flaring reference radius & $R_{0}$ & Explored with MCFOST & 100 AU \\ %
            Flaring reference height & $H_{0}$ & '' & 12, {14} AU {(high/low $f_{\rm UV}$)} \\ %
            Flaring exponent & $\beta$ & '' & 1.15 \\ %
            Disk inner radius & $R_{\rm in}$ & '' & 0.09, 0.05 AU (high/low $f_{\rm UV}$)\\ %
            Minimum dust grain size & $a_{\rm min}$ & '' & 0.05, 0.1 $\rm \mu m$ (high/low $f_{\rm UV}$)\\ %
            Maximum dust grain size & $a_{\rm max}$ & '' & 1 cm \\ %
            Stratification exponent & $s_{\rm set}$ & '' & 0.2, 0.3 (high/low $f_{\rm UV}$) \\ %
            Stratification grain dimension & $a_{\rm set}$ & '' & 0.05, 0.1 $\rm \mu m$ (high/low $f_{\rm UV}$)\\ %
            Inclination & $i$ & '' & $60^{\circ}$ \\ %
            {Disk outer radius} & $R_{\rm out}$ & {''} & {50, 100,} 200 AU \\ %
            Optical extinction & $A_V$ & Free parameter in MCFOST & 1.6 \\ %
            Disk dust mass & $M_{\rm d}$ & '' & $\rm 9 \cdot 10^{-4} \ M_{\odot}$ \\ %
            Cosmic Ray Ionization rate & $\zeta$ & Fixed in \textit{ProDiMo} & $1.7 \cdot 10^{-17}$~s$^{-1}$ \\
            UV excess & $f_{\rm UV}$ & Explored with \textit{ProDiMo} & 0.07, 0.025 \\ %
            PAH abundance &  $f_{\rm PAH}$ & '' & $10^{-2}, 10^{-3}, 10^{-4}$ \\
            Disk gas mass & $M_{\rm g}$ & '' & (9, 4.5, 1.8) $\cdot 10^{-2}$ $\rm M_{\odot}$ \\
            \noalign{\smallskip}
            \hline
            \noalign{\smallskip}
         \end{tabular}
     $$ 

   \end{table*}

\subsubsection{Disk inclination and {extent}} \label{Modeling_inclination}

Guilloteau et al.\ (2011) derived an inclination of $23\pm 14^{\circ}$ from CO sub-millimeter line kinematics. {However, the performed SED modeling indicates that for a disk inclination} $i < 30^{\circ}$, the predicted SED has a NIR bump considerably lower than observed. {On the other hand, if the inclination is} $i > 75^{\circ}$, the photospheric emission is shielded too much by the disk. We found that the SED is well reproduced for $i=60^{\circ}$. However, a fine-tuning of this parameter is hardly feasible and the degeneracy between inclination and inner radius (arising from the analysis of the CO lines, see Sect.\,\ref{Results_diskCO}) remains unsolved (see Appendix \ref{Uncertainties}). 

Assuming an inclination of $60^{\circ}$, the disk inner radius inferred from the profile of the CO ro-vibrational lines is $R_{\rm in}=0.05$ AU (see Eq.\,\ref{Eq_innerradius}). The estimated disk inner radius is at $\simeq$ 6.3 $R_*$, in agreement with estimates of the magnetic truncation radius for typical CTTSs (Shu et al.\ 1994, Donati et al.\ 2008, Long et al.\ 2011). 

{The outer radius of the disk is poorly constrained by the SED. Models with $R_{\rm out}=50$, 100, and 200 AU (all other parameters kept constant) result in the same SED within the photometric error bars (see Sect.\,\ref{Oxygen_line} for a more detailed discussion).}

\subsubsection{Dust grain composition} 
{The 10 $\mu$m silicate feature suggests that sub-$\mu$m size} dust grains are still present in the {surface layer of the} disk (see Sect.\,\ref{Analysis_flaring}). The mineralogy used in our models is amorphous MgFeSiO$_4$ olivine (Dorschner et al.\ 1995). {As the most simple working hypothesis, we assume that the dust is homogeneous in composition throughout the disk. Furthermore, we have} constraints on the gas inner radius $R_{\rm in}$ from the analysis of CO ro-vibrational line profiles ($R_{\rm in}=0.05 \ {\rm AU}$, see Sect.\,\ref{Results_diskCO} and above). If we assume that the dust and gas inner radii are coincident, the dust temperature at that radius has to be below the sublimation temperature. Thus, grains with $a<0.1 \ {\rm \mu m}$ cannot survive at that radius and we use $a_{\rm min}=0.1 \ {\rm \mu m}$. {Another possibility is that a fraction of the CO ro-vibrational line emission originates from gas inside the dust sublimation radius. $a_{\rm max}$ is not well constrained and degenerate with the slope of the grain size distribution. Thus, we {use $a_{\rm max}=1 \ {\rm cm}$, which} is a typical value for disks at this stage. The high $f_{\rm UV}$ model requires a slightly larger $R_{\rm in}$ of 0.09~AU and has a slightly smaller minimum grain size, $a_{\rm min}=0.05 \ \mu$m}.

\subsubsection{{Disk flaring and dust settling}} \label{Flaring}
{The surface mass density of the disk is parametrized as}
\begin{equation}
\Sigma \propto r^{-\epsilon}
\end{equation}
{and the disk scale height as}
\begin{equation}
H(r)=H_0 \cdot \left( \frac{r}{R_0} \right) ^{\beta}
\end{equation}
with {$r$ being the distance from the star} and $H_0$ the disk height at the reference radius $R_0$. According to the qualitative analysis of the IR excess in Sect.\,\ref{Analysis_flaring}, the disk is mildly flared and we thus fixed the flaring angle to be $\beta=1.15$. The best fit resulted in a scale height of $H_0=12 \ \rm AU$ {and $H_0=14 \ \rm AU$ for the high and low $f_{\rm UV}$ respectively at $R_0=100 \ \rm AU$.} Smaller scale heights lead to an underprediction of the FIR fluxes.  

{Dust settling is parametrized assuming that the scale height changes with grain size for grains larger than $a_{\rm set}$} 
\begin{equation}
H(r, a) = H(r) \cdot (a/a_{\rm set})^{-s_{\rm set}/2} 
\end{equation}
The best match of the observed silicate features is found by including dust settling with all particles involved, i.e.\  $a_{\rm set}$= 0.05 $\rm \mu m$, and an exponent $s_{\rm set}$ = 0.2. 
 
\subsection{Gas modeling with \textit{ProDiMo}} \label{Modeling_Prodimo}
\textit{ProDiMo} calculates the chemistry and heating/cooling of the gas self-consistently using a large chemical network of 111 species and 1462 reactions. An extensive list of all heating and cooling processes can be found in Woitke et al.\ (2009, {2012}). In this work, we do not feed the gas temperatures back into the vertical hydrostatic equilibrium, but instead keep the vertical flaring structure given by the MCFOST parametrization found for the best fitting SED model. Using the results from the MCFOST models described in the previous section, we ran a small grid of \textit{ProDiMo} models with different values of UV excess, gas mass and PAH abundance.

\begin{figure*}[htb]
\centering
  \includegraphics[width=7cm]{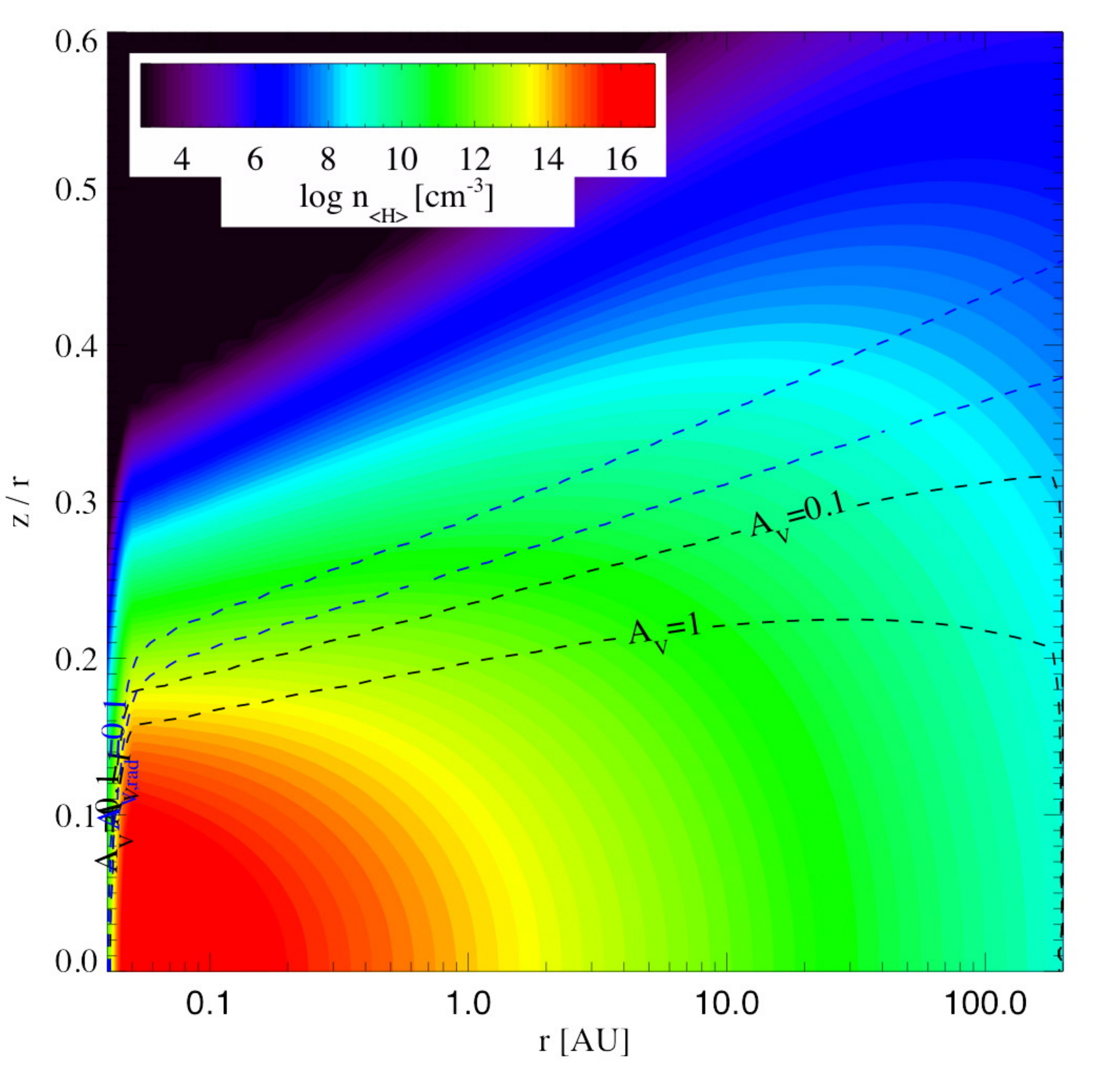}
  \includegraphics[width=7.25cm]{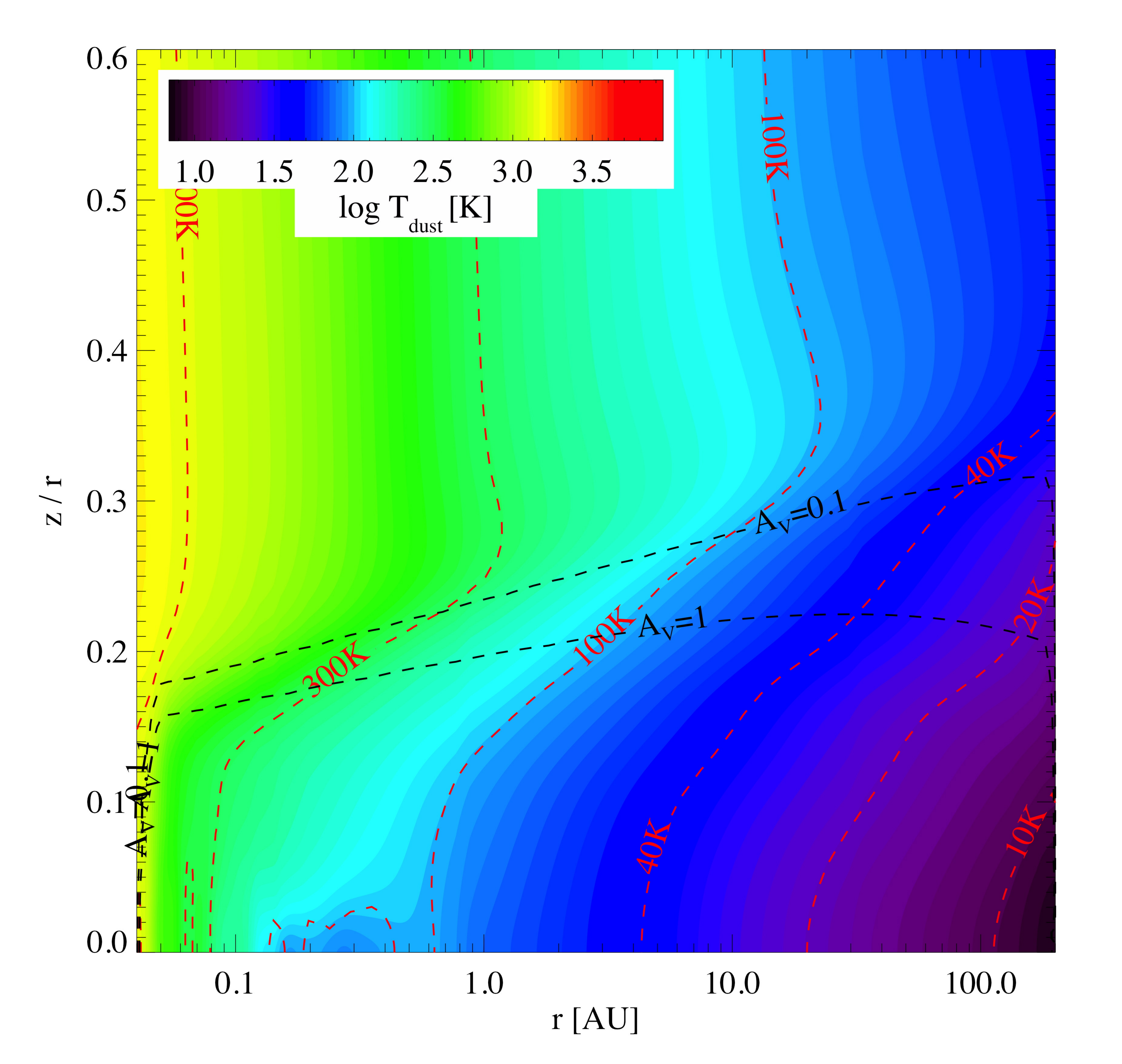}
  \includegraphics[width=7.1cm]{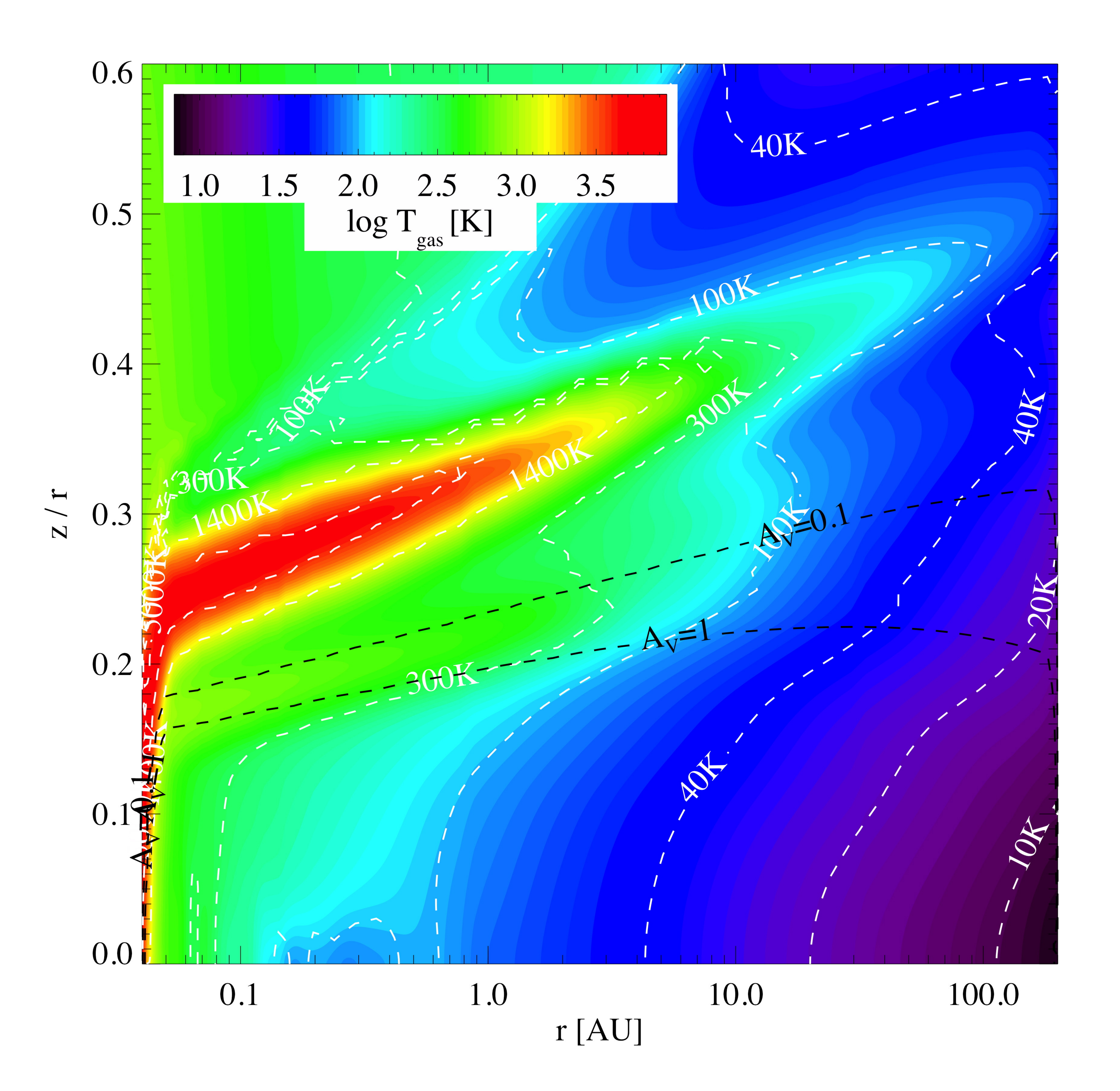}
  \includegraphics[width=7.1cm]{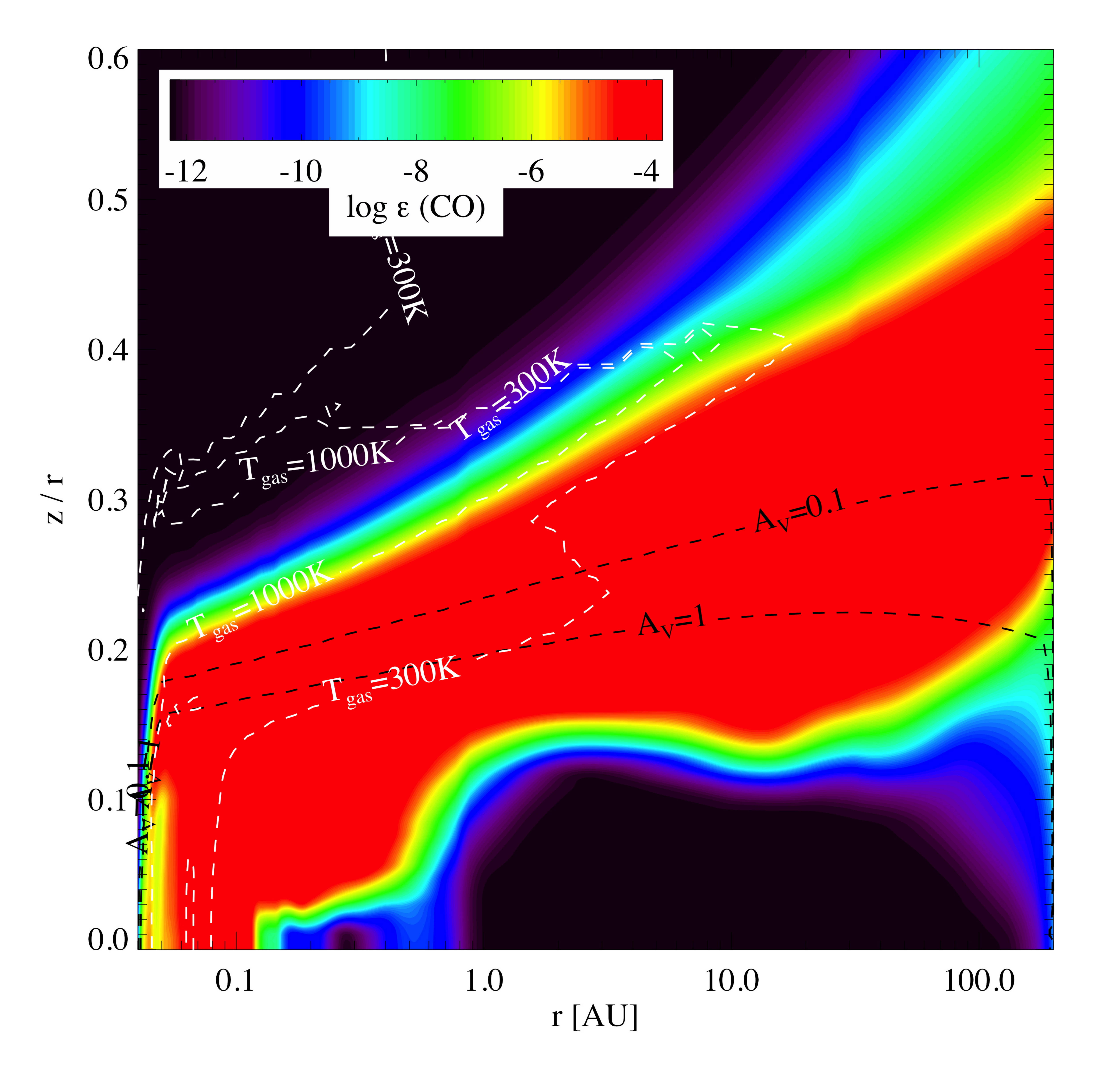}
\caption{From top to bottom, left to right: the total hydrogen number density, the dust temperature, the gas temperature, and the CO abundance distribution in the reference disk model, {namely the case with low UV excess, low gas mass, and low PAH abundance}. The black contours indicate the {total A$_{\rm V}=0.1, 1$ lines}. The white contours indicate the gas temperature, {while the red contours in the dust temperature} plot indicate dust temperature.}
\label{refmodel}
\end{figure*}

{Two UV excess cases were considered (high state, $f_{\rm UV}=0.07$, and low state $f_{\rm UV}=0.025$, see Sect.\,\ref{Modeling_Mcfost})}. We fixed $M_{\rm dust}$ as suggested by MCFOST and explored dust-to-gas mass ratios of 0.01, 0.02, and 0.05, i.e.\ $M_{\rm gas}$=0.090, 0.045, and 0.018 $\rm M_{\odot}$ (hereafter denoted as hGAS, iGAS, and lGAS). Even the most massive model with $M_{\rm gas}$=0.09 $\rm M_{\odot}$ is gravitational stable according to the Toomre criterion (see Eq.\ A.10 of Kamp et al.\ 2011). The abundance of PAHs, $f_{\rm PAH}$, was set to $10^{-2}, 10^{-3}$, and $10^{-4}$ times the one in the ISM (hereafter denoted as hPAH, iPAH, and lPAH). The combination of these three parameters yields a total of 18 disk models.

{The level populations for the line radiative transfer are calculated from statistical equilibrium and escape probability (see Woitke et al.\ 2009, for details). Using these populations, we carry out a detailed line radiative transfer using ray tracing and taking into account the disk rotation and inclination (Woitke et al.\ 2011, Appendix A.7). These detailed radiative transfer fluxes for the eighteen models are listed in Table \ref{COtable}: the $[\rm O\,{\textsc i}] \ 63 \ \mu m$ line and three representative CO ro-vibrational lines, $\nu$=1-0 P4, P36, and $\nu$=2-1 P4.} 

\begin{figure}
\centering
   \includegraphics[width=7cm]{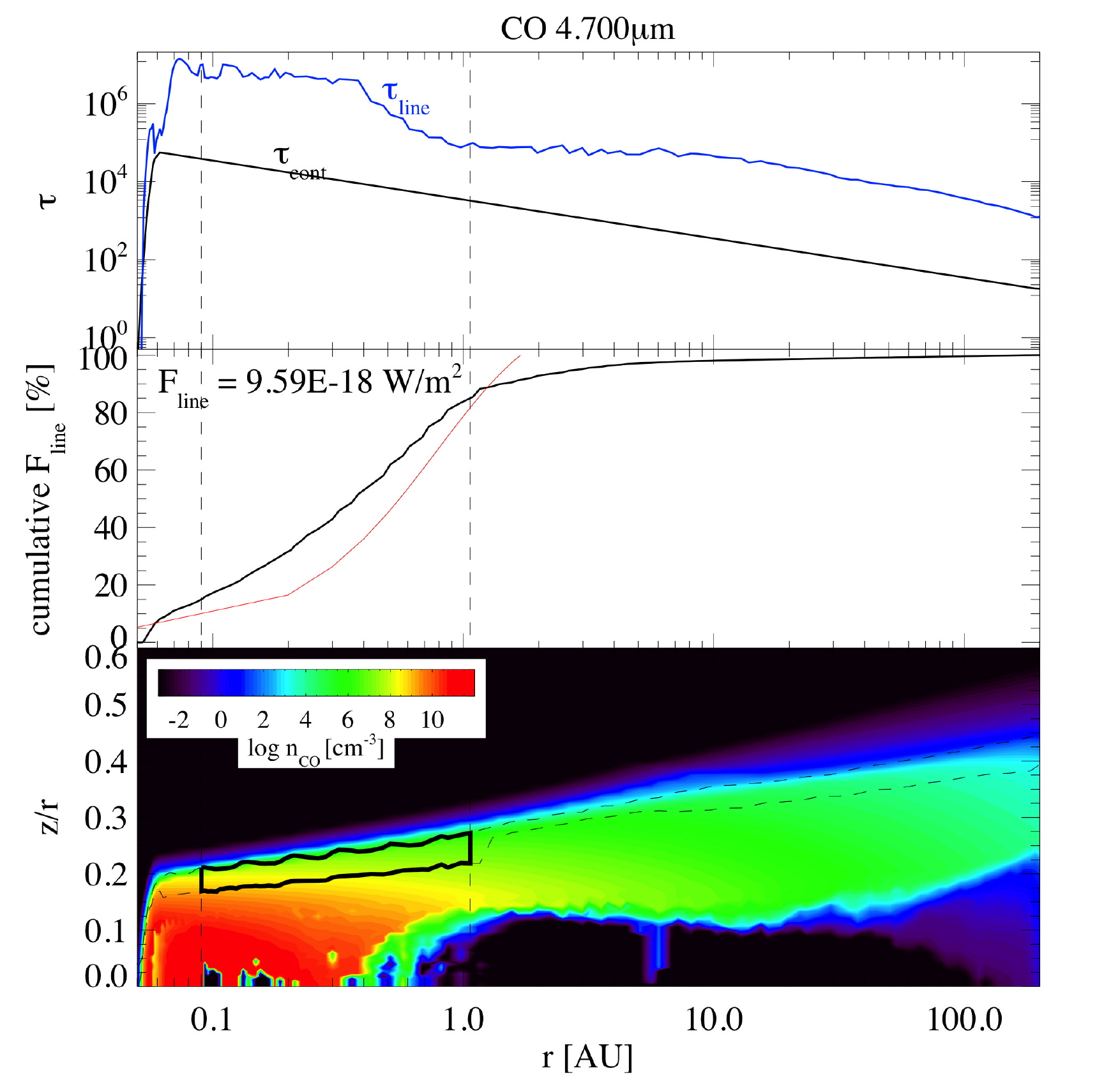}
   \includegraphics[width=7.3cm]{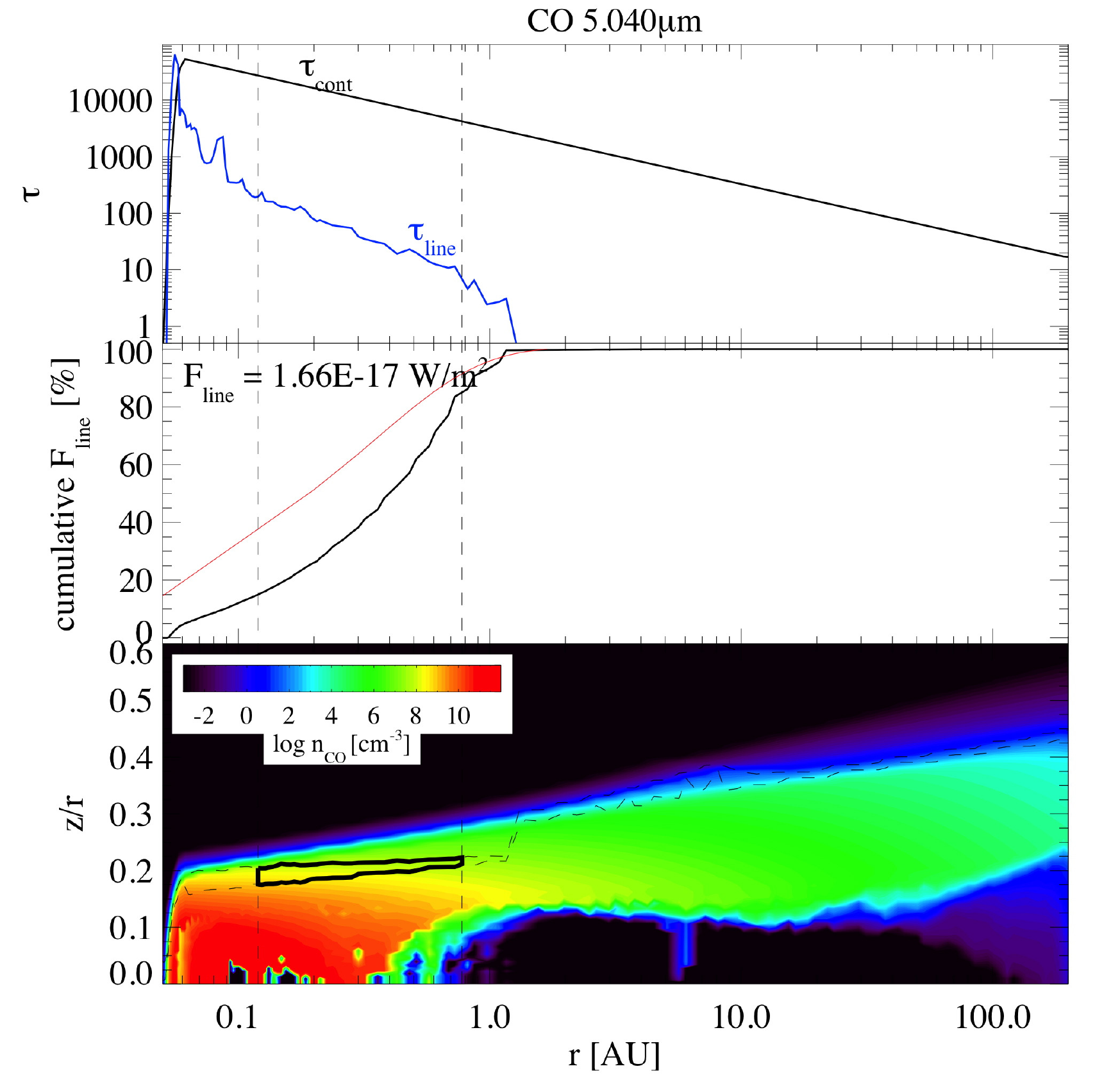}
   \includegraphics[width=7.1cm]{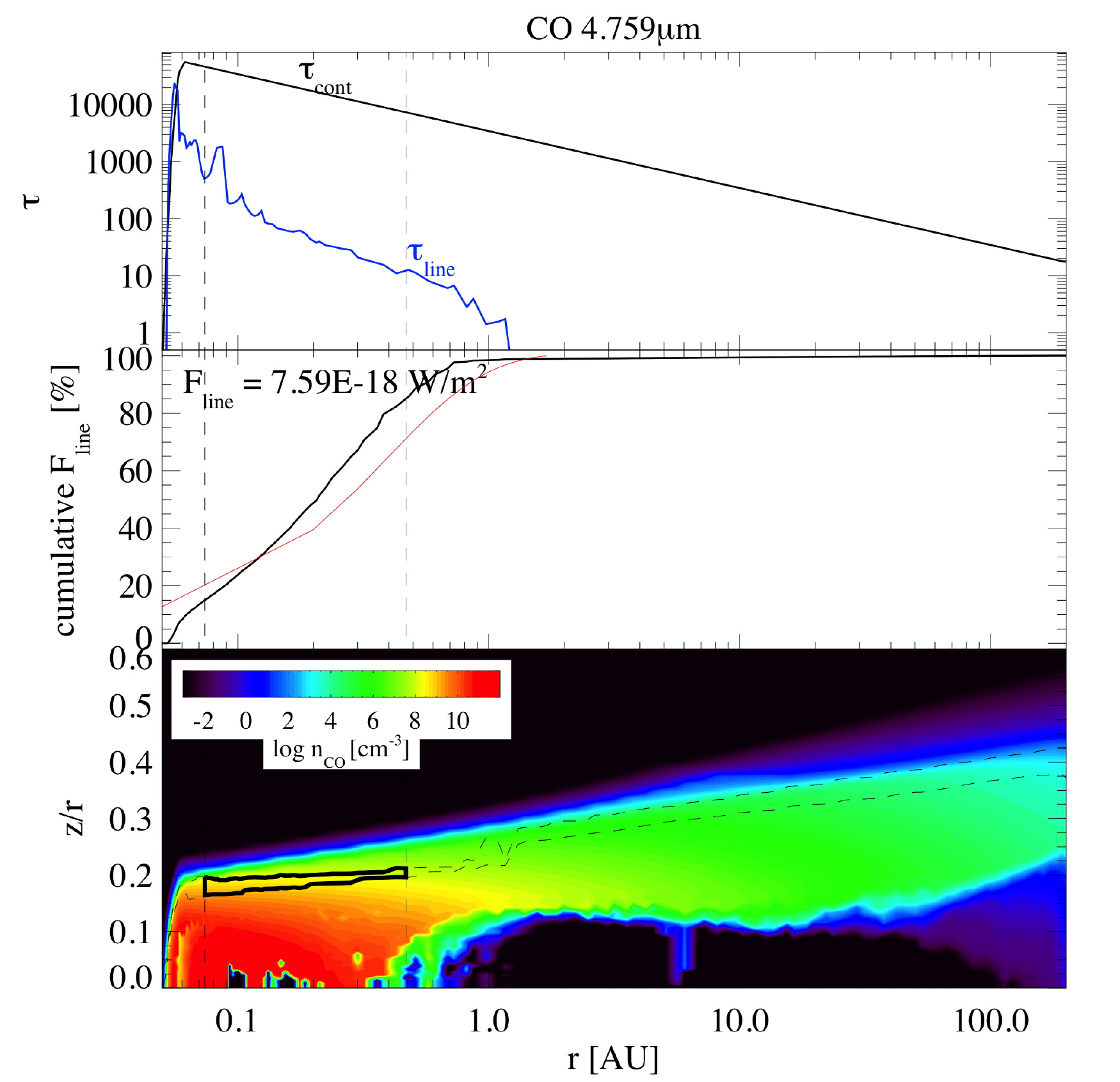}
     \caption{Cumulative flux distribution from vertical escape probability in a \textit{ProDiMo} model with {low UV excess, low gas mass, and low PAH abundance}. The top panel of each box shows the vertical optical depth for the line and continuum. The middle panel of each box shows the cumulative flux from the \textit{ProDiMo} model (black) and best fit slab model (red). The bottom plot of each box shows the CO density distribution in the disk. Outlined with black contours is the region in which radially and vertically between 15 and 85\% of the flux originates. The top box is for the $\nu=1-0$ P4 line, the middle box for the $\nu=1-0$ P36 line, the bottom box for the $\nu=2-1$ P4 line.}
         \label{COP4-cumulative}
   \end{figure}

We chose as a reference model the lUV, {lPAH}, lGAS one. Fig.~\ref{refmodel} illustrates the density and temperature distribution (dust and gas) as well as the CO abundance in that particular model. The dust and gas temperature are well coupled in the region with $A_{\rm V}>1$. The CO abundance reaches a maximum value of $\sim 10^{-4}$ already well above that line and the top CO layer resides at temperatures above 1000~K inside 10~AU. Since the CO {fundamental $\nu=1-0$} ro-vibrational lines are optically thick, they largely originate in this hot surface layer (see also Fig.~\ref{COP4-cumulative}). The main heating process in this region is PAH heating and collisional de-excitation of H$_2$. The main cooling processes are CO rotational and ro-vibrational line cooling as well as water line cooling.

\subsubsection{The CO ro-vibrational lines} 

We use in this study the large CO model molecule compiled by Thi et al.\ (2013) including IR and UV pumping. The model uses 7 vibrational levels of the X$^1\Sigma^+$ and A$^1\Pi$ electronic states and 60 rotational levels within each of them. \textit{ProDiMo} calculates the level populations from statistical equilibrium and performs a detailed line radiative transfer to obtain the emerging CO line fluxes (Woitke et al.\ 2011). This type of thermo-chemical modeling leaves no freedom to adjust CO densities, column densities, densities of collision partners or gas temperatures. 

\begin{figure*}
   \centering
   \includegraphics[width=11.2cm]{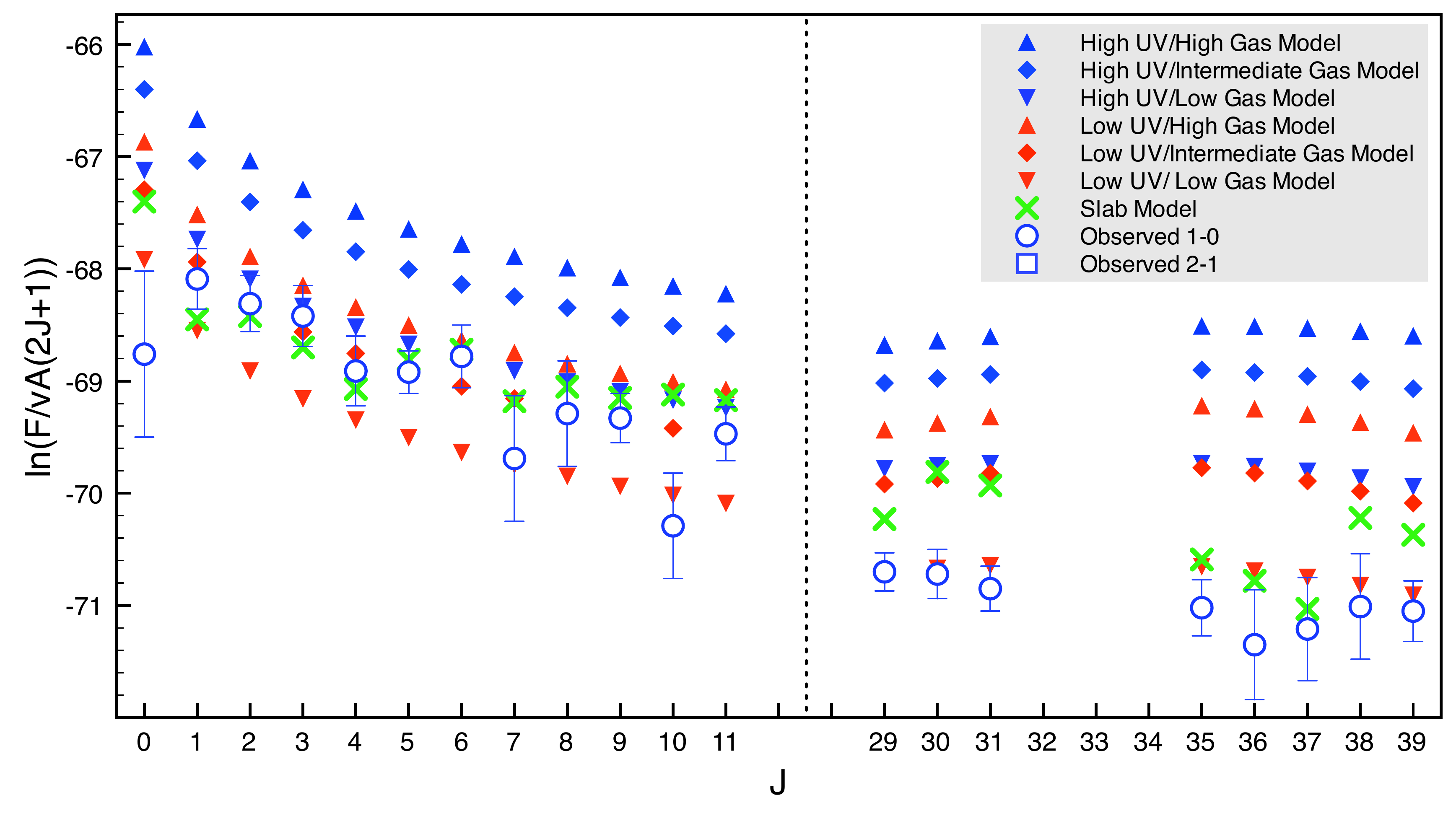}
      \includegraphics[width=5.6cm]{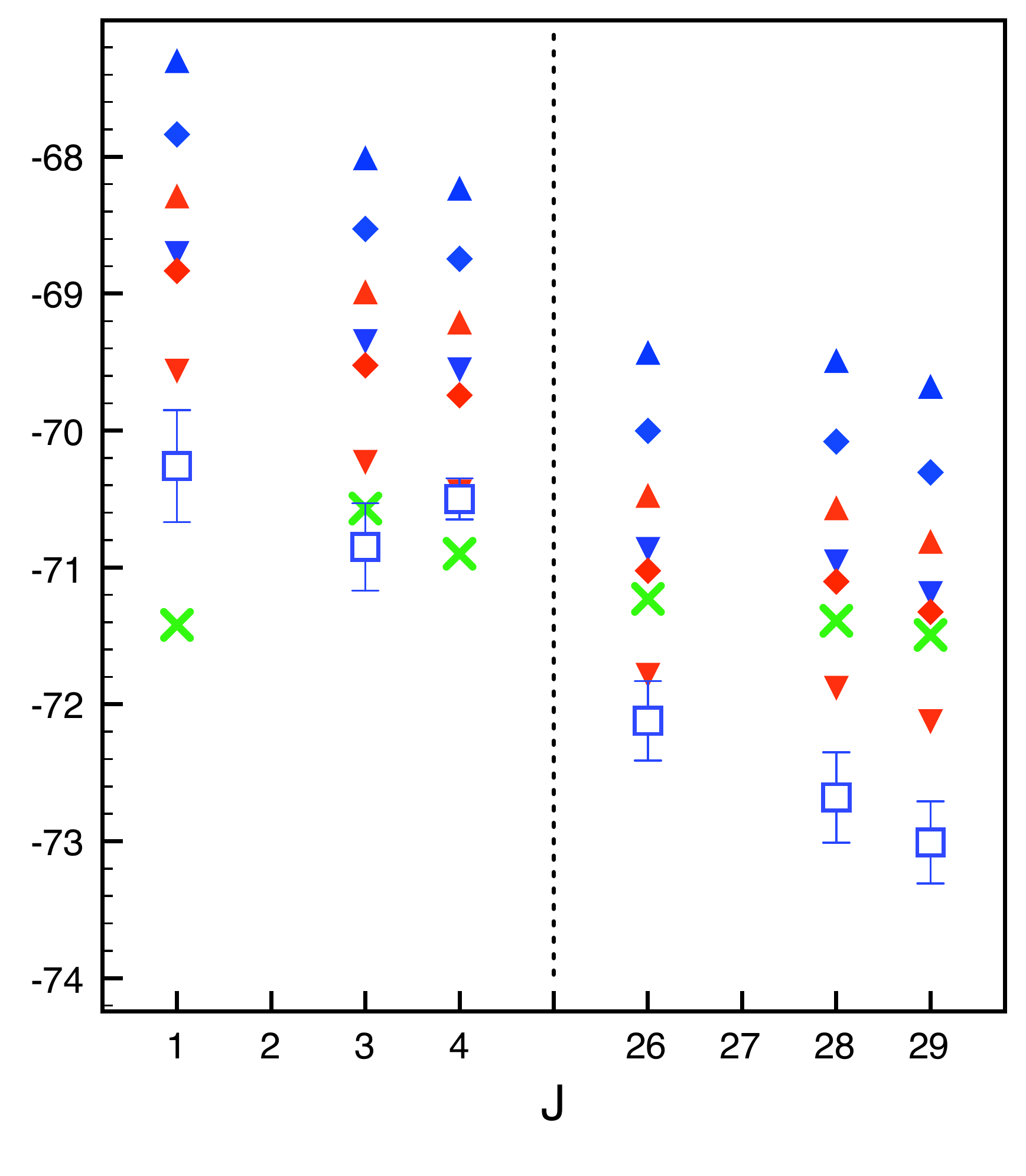}
     \caption{{Rotational diagram of CO ro-vibrational lines observed (left: $\nu = 1-0$, right: $\nu = 2-1$) versus model predicted: slab model described in the text and \textit{ProDiMo} runs with intermediate PAH abundance, with high, intermediate, and low gas mass, and with high and low UV excess. The vertical {dotted} lines indicate discontinuity in the x-axis.}}
         \label{Boltzmann}
   \end{figure*}

The \textit{ProDiMo} models show that CO ro-vibrational line fluxes are very weakly affected by the PAH abundance while they substantially correlate with the gas mass (Table \ref{COtable}). Models with high UV excess generally over-predict the observed fluxes (up to a factor {15}). However, those with low gas {mass} well reproduce or slightly over-predict (by a factor {2}) all $\nu=1-0$ line fluxes. All models with low UV excess agree fairly well with the $\nu=1-0$ and over-predict up to a factor {5} the $\nu=2-1$ line fluxes (see Fig.\ \ref{Boltzmann}). 
  
Alternatively, we also calculated the expected CO line fluxes from a simple line synthesis calculation (see Najita et al.\ 1996 and Brittain et al. 2009 for an extensive description). We assumed that the emission arises from a vertically isothermal slab of gas with constant column density ($N=2 \cdot 10^{18} \ {\rm cm^{-2}}$). {The only collision partner is atomic hydrogen; this is taken as a representative collision partner.} The rotational levels are assumed to be thermalized while vibrational populations are calculated explicitly. From $\chi^2$ minimization, we find a hydrogen volume density profile $n_{\rm H}(r)= 1.5 \cdot 10^{14} \ (r/R_{\rm in, {slab}})^{-2} ~\rm cm^{-3}$ and a gas temperature profile $T(r)= 1200 \ (r/R_{\rm in, slab})^{-0.55}$ K. The inner radius $R_{\rm in, slab}$ is only loosely constrained to $0.1^{+0.1}_{-0.07}$ AU, because the line wings have a rather low S/N. The outer radius is largely unconstrained due to the degeneracy between the surface density of the gas $N$ and the outer radius of the emitting area; $R_{\rm out, slab}$ has to be larger than 0.9 AU. The turbulent line broadening $b$ is found to be 2 km/s, although $N$ and $b$ are degenerate. The gas temperature found at the inner radius is $T_0=1200^{+300}_{-200}$ K. {Integrated line fluxes have been measured from the spectrum generated with the slab model in the same way as for the observed spectra. The model predicts correctly the $\nu = 1 - 0$ lines and the $\nu = 2 - 1$ lines with $T_{\rm ex} \sim 6200$ K. However, the $\nu = 2 - 1$ lines at higher energy are over-predicted by a factor 5 (see Fig.\,\ref{Boltzmann}). This could indicate that the gas is more diffuse (lower volume density), that the line flux declines steeper with distance from the star (steeper density power-law), or that some additional non-LTE effects are still missing in the slab model. }

Fig.\,\ref{COP4-cumulative} shows the cumulative flux distribution {from simple vertical escape probability} in the lUV/lPAH/lGAS \textit{ProDiMo} model for the three representative lines, $\nu$=1-0 P4, $\nu$=2-1 P4, and $\nu$=1-0 P36. The CO lines are sub-thermally excited; {in the hUV/lPAH/lGAS model, the three lines studied here as representative lines are a factor 2-3 lower than their respective LTE values}. The agreement between \textit{ProDiMo} models and the more simple slab models on how the flux is building up as a function of radius is very good (see comparison in Fig.\,\ref{COP4-cumulative}). The \textit{ProDiMo} models also show a similar temperature of $\sim 1000$ K at the inner radius of the CO ro-vibrational line emitting region (compared to 1200 K from best fit slab model).

{It turns out that the CO line fluxes depend on gas mass, UV excess, and PAH abundance (see Table \ref{COtable}). Hence, the lines cannot be used to directly constrain the disk gas mass.} The apparently better fit of the simple CO line modeling could be largely due to the fact that parameters such as CO column density, gas temperature and collision partner density can be varied independently without imposing a self-consistent disk structure. This means that CO in a slab model can for example {exist} in regions where it would be photodissociated in a thermo-chemical disk model, thus allowing UV fluorescence to affect the CO ro-vibrational lines much stronger in the former case. {Another possibility could be that part of the CO ro-vibrational emission originates from gas inside the dust sublimation radius of our models.} At this stage, a further analysis of these CO ro-vibrational lines is largely limited by the observations, which have limited spectral resolution and suffer from a low signal-to-noise and telluric contamination (see Sect.\,\ref{Observations_Keck}).

\subsubsection{The [O {\sc i}] 63 $\mu$m line} 

As shown in Table \ref{COtable}, the [O\,{\textsc i}]\,63\,$\mu$m line {flux} is affected by the {UV excess}, the gas mass, and the PAH abundance. Differences of a factor two are found between the high and low UV excess models. The dependence on the gas mass stems from the fact that the gas temperature changes with disk mass and, in turn, affects the line flux. The fact that the flux increases with PAH abundance is explained by the increasing photoelectric heating of the gas in the upper disk layer (Jonkheid et al.\ 2004). Thus, gas mass, UV excess and PAH abundance are to some degree degenerate in the prediction of the [O\,{\textsc i}] line flux. 

\begin{table}
      \caption[]{Fluxes of the [O\,{\textsc i}]\,63\,$\mu$m line and of three representative CO ro-vibrational lines as predicted by the grid of \textit{ProDiMo} models {using detailed line radiative transfer}, the slab model described in the text, and as observed.}
         \label{COtable}
     $$ 
         \begin{tabular}{lcccc}
            \hline
            \hline
            \noalign{\smallskip}
            Model & \multicolumn{3}{c}{Flux ($\rm 10^{-17} \ W/m^2$)}\\
            & [O\,{\textsc i}] & CO& CO& CO \\
            UV / PAH / GAS & 63 $\mu$m & 1-0 P4 & 2-1 P4 & 1-0 P36 \\
            \hline
            \noalign{\smallskip}
           h / h / h & 30.4 & 6.51 & 5.07 & 10.4 \\
ÊÊÊÊÊÊÊÊÊÊÊh / i / h & 21.7 & 5.10 & 4.79 & 10.3 \\
ÊÊÊÊÊÊÊÊÊÊÊh / l / h & 20.9 & 4.95 & 4.77 & 10.3 \\
ÊÊÊÊÊÊÊÊÊÊÊh / h / i & 24.7 & 4.35 & 2.99 & 7.05 \\
ÊÊÊÊÊÊÊÊÊÊÊh / i / i & 18.3 & 3.55 & 2.86 & 6.95 \\
ÊÊÊÊÊÊÊÊÊÊÊh / l / i & 17.7 & 3.41 & 2.84 & 6.59 \\
ÊÊÊÊÊÊÊÊÊÊÊh / h / l & 17.5 & 2.11 & 1.31 & 3.09 \\
ÊÊÊÊÊÊÊÊÊÊÊh / i / l & 13.4 & 1.81 & 1.26 & 3.02 \\
ÊÊÊÊÊÊÊÊÊÊÊh / l / l & 12.9 & 1.77 & 1.25 & 3.02 \\
           \noalign{\smallskip}
            l / h / h & 15.6 & 2.97 & 2.03 & 5.31 \\
            l / i / h & 11.3 & 2.17 & 1.80 & 5.05 \\
            l / l / h & 10.8 & 2.08 & 1.78 & 5.03 \\
            l / h / i & 13.1 & 1.94 & 1.15 & 3.04 \\
            l / i / i & 9.92 & 1.44 & 1.06 & 2.91 \\
            l / l / i & 9.59 & 1.38 & 1.05 & 2.90 \\
            l / h / l & 10.1 & 1.03 & 0.55 & 1.26 \\
            l / i / l & 7.99 & 0.79 & 0.52 & 1.21 \\
            l / l / l & 7.77 & 0.77 & 0.52 & 1.20 \\
            \hline
            Slab model & - & 1.25 & 0.40 & 1.31 \\
            \hline
            \noalign{\smallskip}
            Observed & 1.6$\pm 0.5$ & {1.65$\pm 0.38$ } & {0.28$\pm 0.10$} & {0.83$\pm 0.08$} \\
            \noalign{\smallskip}
            \hline
             \end{tabular}
     $$ 

   \end{table}

\section{Discussion} \label{Discussion}
In this section, we discuss the source variability and the results of our detailed disk modeling in the context of the available observational data. 

\subsection{Source variability} \label{Variability}
Young circumstellar systems are often highly variable objects (see e.g.\ Bouvier et al.\ 1993). Flux variability up to a factor $\sim 3$ has been measured for FT Tau in the V band on a timescale of five years (ASAS, Pojmanski 2002). Furthermore, flux variability up to a factor $\sim$ 2.8, 2.1, and 1.3 has been measured in B, V, and I band respectively on a timescale of 45 days (Fern\'{a}ndez, private comm.) with the SITe CCD attached to the 1.23 m telescope of the Calar Alto Observatory (Almer\'ia, Spain). The analysis of this variability revealed that it cannot be reproduced by cold spots in the stellar photosphere because the amplitude of the variations observed in the V band is too large with respect to variations in the I band. On the contrary, these amplitudes are well matched by photospheric hot spots (from 5000 to 6600 K) if we assume the effective temperature of M3 stars. The B band shows an amplitude slightly smaller than expected for those hot spots but this can be explained by the presence of a hot continuum in addition to the pure photospheric emission. 

\begin{figure*}
   \centering
   \includegraphics[width=16cm]{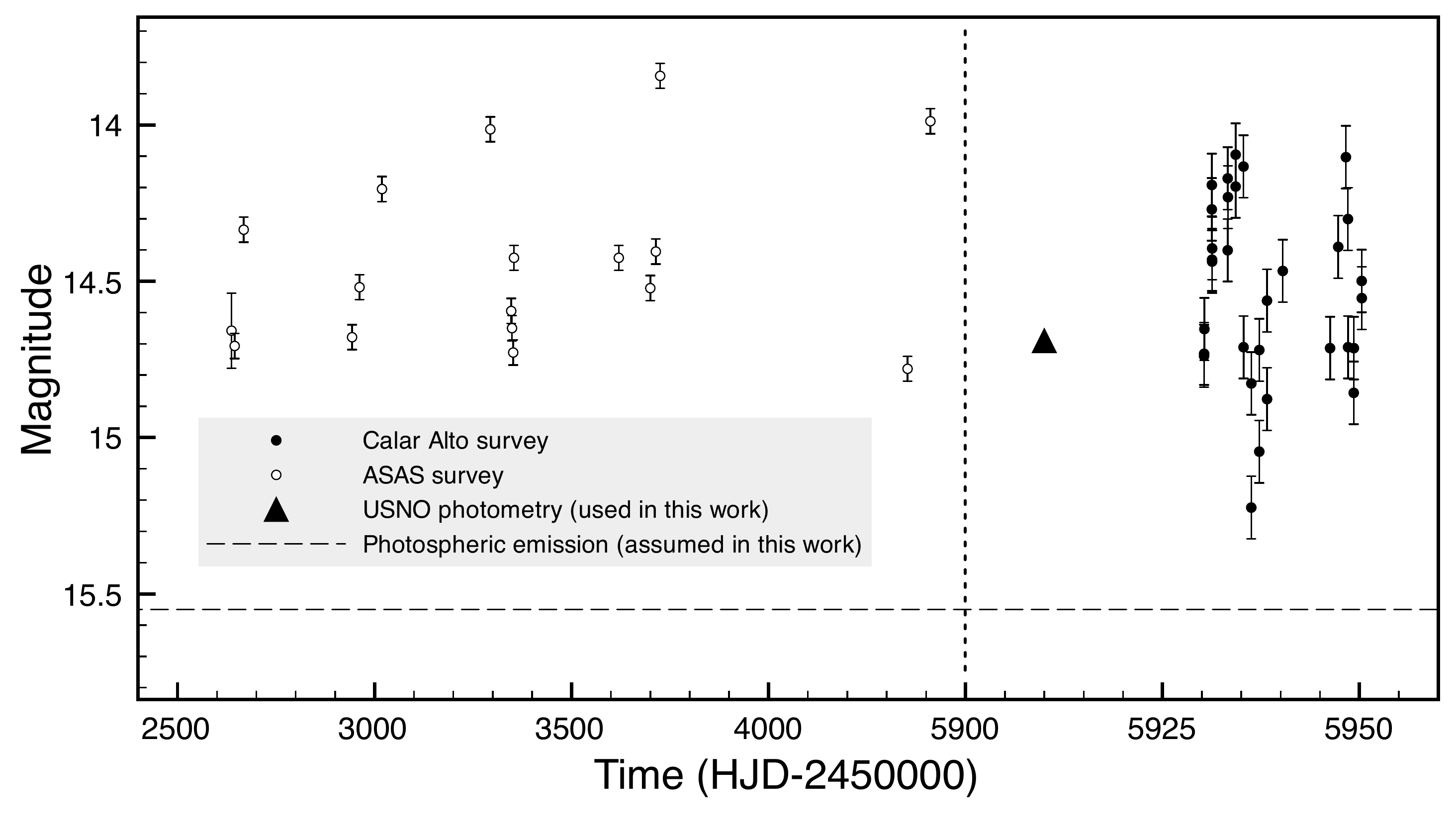}
     \caption{ASAS, Calar Alto, and USNO photometric measurements of FT Tau in the V band. The dashed line indicates the reddened photospheric magnitude assumed in our analysis. The vertical line indicates a gap in the x-scale. The position in time of the USNO photometry is arbitrary.}
         \label{V_var}
   \end{figure*}

Given this, the brightness at the minimum of the light curves provides an upper limit to the photospheric brightness of the star. As we see from Fig.\,\ref{V_var}, the reddened photospheric emission in the V band assumed in Sect.\,\ref{Results_stellar} is lower than measurements, from either ASAS or Calar Alto surveys. This indicates that the extinction cannot be much lower than estimated in Sect.\,\ref{Results_stellar}, because this would increase the reddened photospheric emission of the model to values higher than the observed one. The USNO V band photometry used to flux-calibrate the TNG spectrum (and, thus, to estimate the mass accretion rate, see Sect.\,\ref{Observations_Tng} and \ref{Results_accretion}) turns out to be an average value of all measurements (see Fig.\,\ref{V_var}).  

In order to address the origin of the observed variability, a time-dependent study of optical/NIR emission lines is necessary. Any relation between these lines and contemporary observations of optical photometric variations can clarify if and how much of the observed variability is due to the accretion process. In addition, we must be careful in the interpretation of line emission from the disk surface especially if these lines result from UV pumping by stellar radiation.

\subsection{{The CO ro-vibrational lines}}
{The CO ro-vibrational lines are very sensitive to the extent of the hot gas surface layer. The observations clearly indicate that the lines are
typically very wide (HWZI $\simeq 65$ km/s). In the models the CO ro-vibrational lines predominantly arise from this hot surface layer (Sect.\,\ref{Modeling_Prodimo}). The UV radiation field affects the extent of this hot surface and it can change due to the particular choice of the dust opacities, the scale height of the disk and the flaring. The quality of the available Keck CO ro-vibrational line profiles is not good enough to derive the extent of the hot surface layer directly from their shape. In case of exquisite data quality, this can be done as shown by Goto et al.\ (2012) for the example of HD100546, a Herbig Ae star. So, as new data will become available, these parameters should be refined keeping the constraints on the SED.}

\subsection{The [O {\sc i}] 63 $\mu$m line} \label{Oxygen_line}
All models presented here over-predict the [O {\sc i}] 63 $\mu$m line. Since the line is optically thick, its flux is mostly affected by the gas temperature in the emitting region and the total emitting surface area. {The models indicate that} the [O {\sc i}] 63 $\mu$m line typically originates between $\sim 10$ and 200 AU. Roughly 15\% of the total line flux builds up between 100 and 200 AU. {In order to understand the dependence of the predicted [O {\sc i}] line flux on the adopted disk size, we calculated models with different outer radii (50, 100, and 200 AU, see Table \ref{Model}). We find that the line flux decreases by only a factor 4 for the smallest disk size. This is due to the fact that the smaller emitting area is partially compensated by the higher gas temperature of the emitting region. At the same time, the CO ro-vibrational lines do not change within the modeling uncertainties. Hence, the observed emission lines do not allow us to put any stronger constraint on the size of the gaseous disk.} 

Guilloteau et al.\ (2013) derive from IRAM 30-m ${\rm CN\ N=2-1}$ observations an outer radius of 310 AU. Previously, a simple power law disk model fit to 1.3 mm and 2.7 mm interferometric continuum data yielded $R_{\rm out} = 57$ AU (Guilloteau et al.\ 2011). However, the disk is barely resolved and {better interferometric images} at shorter wavelength are required to measure $R_{\rm out}$ for the dust; at the same time, interferometric line data e.g.\ for CO isotopologues {are} required to obtain a reliable outer gas radius. Previous work shows that gas and dust outer radii at submm wavelength can actually differ (e.g.\ Isella et al.\ 2007, Andrews et al.\ 2012). {Given the existing uncertainty on the estimate of $R_{\rm out}$, models with different gas and dust outer radii have to await better observational data.}

In the region where the [O {\sc i}] 63 $\mu$m line emits, photoelectric heating is one of the dominant heating processes. Since the PAH features are not observed in the Spitzer spectra, their abundance can be arbitrarily low. Supressing the PAH abundance even below the lowest value in the grid, $f_{\rm PAH}=10^{-4}$, does not affect the [O {\sc i}] line flux anymore. A lower disk gas mass shifts the line forming region to lower depth in the disk, thus making the line flux weaker.

\section{Summary}
We have performed analysis and modeling of the SED and emission lines of the TTS FT Tau to fully characterize the stellar, disk, and accretion properties. We reduced and analysed five spectra from optical to FIR wavelengths, taken with the optical Telescopio Nazionale Galileo, the NIR William Herschel Telescope, the NIR Nordic Optical Telescope, the high-resolution NIR Keck Observatory, and the FIR {\it Herschel} Space Telescope. Additional data were retrieved from the literature and consist of a MIR Spitzer Space Telescope spectrum and of 44 photometric measurements from optical to radio wavelengths. We studied a set of models of the source generated by means of the radiative transfer code MCFOST and the thermo-chemical disk modeling code \textit{ProDiMo}. 

We found that FT Tau is a {low-mass ($\rm 0.3 \pm 0.1 \ M_{\odot}$) and -luminosity ($\rm 0.35 \pm 0.09 \ L_{\odot}$) M3} star showing very high variability (probably due to photospheric hot spots). The estimated properties are typical of young unevolved systems in the Taurus star forming region {with ages of roughly 1 Myr}. The optical extinction ($A_V=1.8$) is {also} within the range of typical values in Taurus. 

The inner radius of the circumstellar disk is small ($0.05 ^{+0.04}_{-0.02} \ {\rm AU}$) indicating an early stage of internal disk dissipation. This is in agreement with the fact that the {source is strongly accreting. In fact, the derived mass accretion rate ($(3.1 \pm 0.8) \cdot 10^{-8}\ \rm M_{\odot}/yr$) is a high value for M-stars, since this corresponds to $L_{\rm acc} \sim 0.4 \ L_*$}. The {ratio of mass outflow to mass accretion rate is lower than $0.03$, in agreement with typical observed values for TTSs (Hartigan et al.\ 1995).} The disk is quite massive ($\sim 0.02 \rm \ M_{\odot}$) with respect to the stellar mass (mass ratio $\sim 0.06$). {Small silicate grains are still present in the disk surface even though at lower abundance due to some degree of settling. Winds and radiation pressure seem to be inefficient at removing this small grain population. These findings are consistent with the apparent primordial nature of this disk (e.g.\ no gaps, holes).} The PAH abundance is inferred to be extremely low ($\sim 10^{-4}$ times that in the ISM). 

From this work it is clear that FT Tau can be considered as benchmark for primordial disks in the Taurus molecular cloud with high mass accretion rate, {high gas content}, and typical disk size. It is an interesting target for follow-up chemical studies as well as for informing surveys of star forming regions on more prototypical objects to expect.

\begin{acknowledgements}
We acknowledge the referee for valuable comments that considerably improved the paper. We thank Matilde Fern\'{a}ndez and Victor Terr\'{o}n for observing FT Tau and reducing the data at the Calar Alto Observatory. We really appreciate the helpful discussion about the origin of the variability. We also gratefully thank Gwendolyn Meeus for her work in acquiring data with the TNG and Ilaria Pascucci and Veronica Roccatagliata for reducing the data from Spitzer. This work is supported by the Swiss National Science Foundation. LP acknowledges the funding from the FP7 Intra-European Marie Curie Fellowship (PIEF-GA-2009-253896). IK, WFT, FM, and PW acknowledge funding from an NWO MEERVOUD grant and from the EU FP7- 2011 under Grant Agreement nr.\ 284405. FM acknowledges support from the Millennium Science Initiative (Chilean Ministry of Economy), through grant ÒNucleus P10-022-FÓ. I.\ Pascucci acknowledges NASA/ADP Grant NNX10AD62G. This research has made use of the SIMBAD database, operated at CDS, Strasbourg, France.
\end{acknowledgements}

\begin{appendix}

\section{Uncertainties of the analysis} \label{Uncertainties}

In this appendix we discuss the limitations of our analysis due to non-simultaneous observations and quantify thoroughly the uncertainties on the inferred results.

The mentioned stellar variability does not have strong impact on the determination of the stellar properties because it does not affect significantly the shape of the optical spectrum (used to determine the spectral type) and the J and H band fluxes (used to estimate luminosity and radius).

\begin{figure}
   \includegraphics[width=8.5cm]{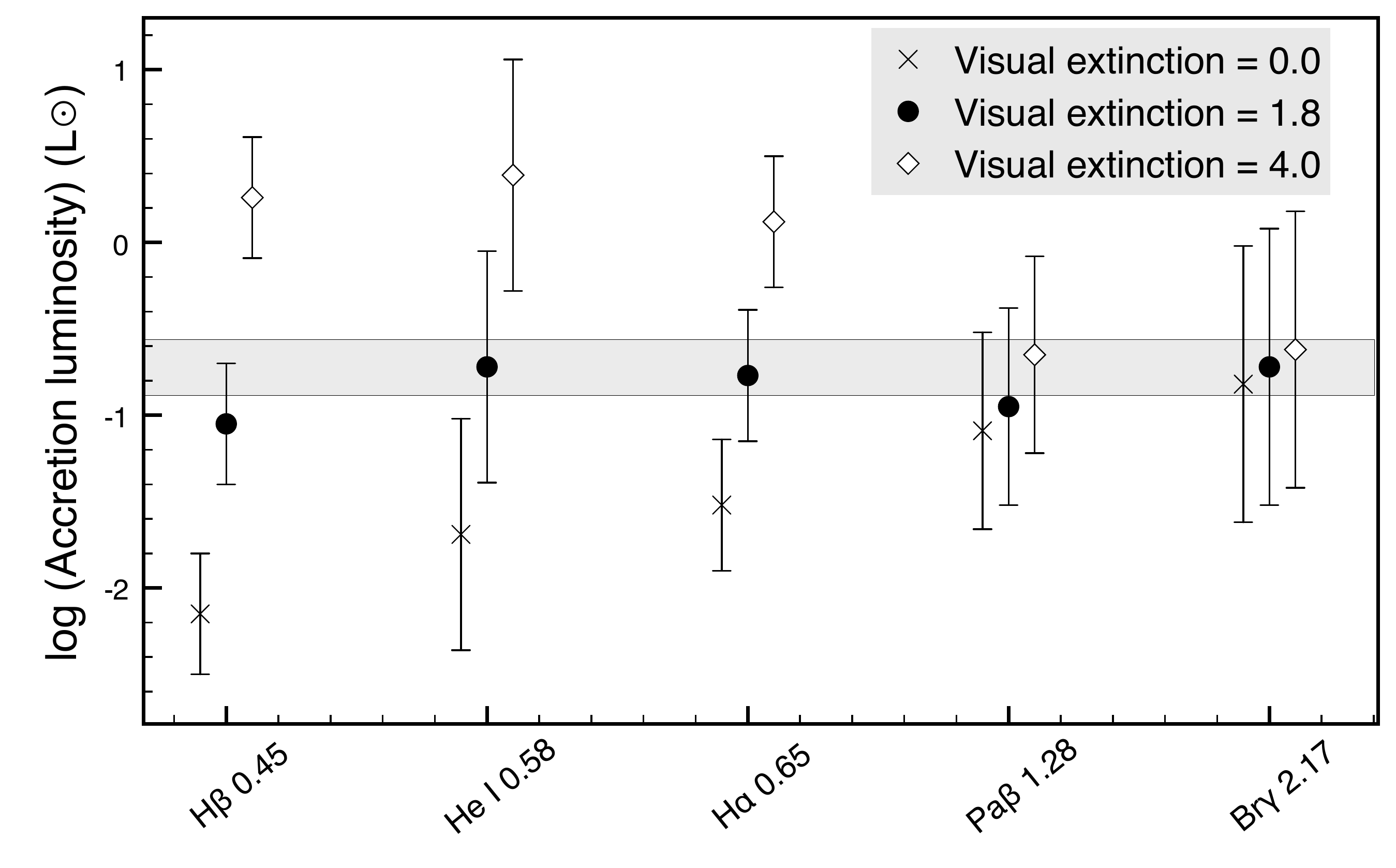}
     \caption{Accretion luminosity estimated from the luminosity of emission lines at optical to NIR wavelengths (x-axis) for different $A_V$ values. The grey stripe indicates the range of values inferred from the Br$\gamma$ line, which is the least affected by extinction. It is clear the trend with wavelength for high and no extinction. Slight displacement between points has been put for a better visualization.}
         \label{Accr_estimates}
   \end{figure}

On the contrary, the estimate of the visual extinction is affected by large uncertainties. We firstly remark that the use of the (J-H) color as tracer of the extinction relies on the assumption that the observed flux at those wavelengths is entirely emitted by the stellar photosphere. Secondly, the determination of the optical extinction $A_V$ is pretty sensitive to the surface gravity of the assumed model. By varying the stellar radius or mass by 30\%, we obtain $A_V$ values between 1.2 and 2.5. This may add a further factor 15\% uncertainty to the estimates of stellar properties. However, the fact that the accretion luminosity values estimated by using different tracers from 0.45 and 2.17 $\mu $m does not show a dependence with the wavelength (see Fig.\ \ref{Accr_estimates}) is a strong sanity check for the determination of $A_V$. The fact that we find the same $A_V$ values by using two independent methods (the observed colors, Sect.\,\ref{Results_stellar}, and the modeling approach, Sect.\,\ref{Modeling_Mcfost}) further reinforces our result. The large difference between our estimate of the stellar luminosity and the result from Rebull et al.\ (2010) (see Table \ref{Properties}) is due to the determination of $A_V$ which is in turn due to the assumed surface gravity.

The spectral type-$T_{\rm eff}$ relation can actually introduce an additional error. Differences up to some hundreds of Kelvin {arise} for M-type stars among different works (see e.g.\ Da Rio et al.\ 2010). Finally, further uncertainty in the determination of the stellar properties is provided by the PMS star tracks adopted to infer the stellar mass and age. {Hartmann (2001) suggested that the age spread inferred for TTSs in Taurus may exclusively be due to uncertainties towards individual members.}

In Sect.\,\ref{Results_diskCO} we estimated the disk inner radius by measuring the width of the CO ro-vibrational lines. The largest uncertainty in the determination of $R_{in}$ is set by the adopted inclination. The width of the CO lines is equally reproduced by configurations with ($i$ : $R_{\rm in}$) = ($60^{\circ}$ : 0.05 AU), ($45^{\circ}$ : 0.03 AU), and ($30^{\circ}$ : 0.02 AU).
 
The estimates of the mass accretion and outflow rate may be affected by variability, since the optical and NIR spectra used to measure the line luminosities were flux-calibrated by using non-simultaneous photometry. This is particularly true for estimates based on optical lines (optical flux variability $\sim$ 2.1, see Sect.\,\ref{Variability}). The  variability implies uncertainties on the estimates of the accretion luminosity in addition to the scattering of the empirical correlations employed to derive $L_{\rm acc}$ (Sect.\,\ref{Results_accretion}). The lowest and the highest estimates for  $L_{\rm acc}$ have been found by means of emission lines from the same spectrum (thus taken simultaneously, see Table \ref{Accretion}). This is indicating that the scattering of the empirical relations might play the major source of uncertainty on the accretion luminosity.

\end{appendix}

\end{document}